\documentclass[english,aps,prl,twocolumn,amsmath,amssymb,letterpaper,showpacs,superscriptaddress]{revtex4-1}
\usepackage{amsmath}
\usepackage{amssymb}
\usepackage{nicefrac}
\usepackage{graphicx}
\usepackage{pdfpages}	% Packages needed
\usepackage{pgffor}		% to include supplement.pdf
\usepackage[unicode=true,pdfusetitle,
 bookmarks=true,bookmarksnumbered=false,bookmarksopen=false,
 breaklinks=false,pdfborder={0 0 1},backref=false,colorlinks=false]
 {hyperref}

\makeatletter
\@ifundefined{textcolor}{}
{%
 \definecolor{BLACK}{gray}{0}
 \definecolor{WHITE}{gray}{1}
 \definecolor{RED}{rgb}{1,0,0}
 \definecolor{GREEN}{rgb}{0,1,0}
 \definecolor{BLUE}{rgb}{0,0,1}
 \definecolor{CYAN}{cmyk}{1,0,0,0}
 \definecolor{MAGENTA}{cmyk}{0,1,0,0}
 \definecolor{YELLOW}{cmyk}{0,0,1,0}
}

\date{\today}

\makeatother

\begin{document}

\title{Microwave signatures of Majorana states in a topological Josephson
junction}

\author{Jukka I. V\"ayrynen}

\affiliation{Department of Physics, Yale University, New Haven, CT 06520, USA}

\author{Gianluca Rastelli}

\affiliation{Zukunftskolleg, Universit\"at Konstanz, D-78457 Konstanz, Germany}
\affiliation{Fachbereich Physik, Universit\"at Konstanz, D-78457 Konstanz, Germany}

\author{Wolfgang Belzig}

\affiliation{Fachbereich Physik, Universit\"at Konstanz, D-78457 Konstanz, Germany}

\author{Leonid I. Glazman}

\affiliation{Department of Physics, Yale University, New Haven, CT 06520, USA}

\begin{abstract}
We find the admittance of a topological Josephson junction $Y(\omega,\varphi_0, T)$ as a function of frequency $\omega$,  the static phase bias $\varphi_0$ applied to the superconducting leads, and temperature $T$. The dissipative part of $Y$ allows for spectroscopy of the sub-gap states in the junction. The resonant frequencies  $\omega_{\mathcal{M}, n} (\varphi_0)$ for transitions involving the Majorana ($\mathcal{M}$) doublet exhibit characteristic kinks in the $\varphi_0$-dependence at $\varphi_0=\pi$. The kinks -- associated with  decoupled Majorana states -- remain sharp and the corresponding spectroscopic lines are bright at any temperature, as long as the leads are superconducting. The developed theory may help extracting quantitative information about Majorana states from microwave spectroscopy.
\end{abstract}

\maketitle

The interest in the condensed matter realizations of Majorana states is fueled by the promise of  topologically-protected quantum computing~\cite{Kitaev20032,2008RvMP...80.1083N,2012RPPh...75g6501A,Beenakker13}. While the latter requires the ability to braid the states, the current experimental effort~\cite{mourik_signatures_2012,rokhinson_fractional_2012,Deng2012,2012NatPh...8..887D,2011PhRvB..84v4521K,Kurter13,hart_induced_2013,Pribiag14} focuses on indications of the Majorana states' presence in various implementations of topologically-nontrivial superconductors. The majority of experiments use the dc electron transport spectroscopy and aim at detecting a zero-bias conductance peak associated with tunneling into a Majorana state~\cite{law2009,flensberg_tunneling_2010,Sau2010a}
 or the $4\pi$-periodic phase dependence of the two ``Majorana branches'' formed by an occupied and unoccupied  Majorana doublet~\cite{kwon_fractional_2004,rokhinson_fractional_2012,fu_josephson_2009}. In the former case, some additional checks are necessary~\cite{mourik_signatures_2012} to exclude other sources ({\it e.g.}, Kondo effect) of the zero-bias anomaly~\cite{Churchill13,2012NatPh...8..887D,Van_Harlingen13}. Attempts to observe the unusual phase dependence rely on the absence of inter-branch relaxation; this is hard to enforce, especially over an extended time period required in the interference experiments~\cite{Kurter13,hart_induced_2013,Pribiag14,badiane_ac_2013}, or at higher bias voltage needed for observation of multiple Shapiro steps~\cite{rokhinson_fractional_2012}.

Limitations of the techniques implemented to-date give an incentive to search for alternatives. Our theory elucidates the manifestations of the Majorana states in the microwave spectrum of a topological Josephson junction. Spectroscopy of the Andreev ({\it i.e.}, sub-gap) states was performed recently in experiments with conventional metallic break junctions~\cite{bretheau2013a,Bretheau2013b}. The experiments did detect the transitions from an Andreev level to the continuum of quasiparticle states~\cite{Bretheau2013b}, and the transitions between the two discrete Andreev levels~\cite{bretheau2013a}. The latter result in a narrow bright line in the spectrum, especially attractive for spectroscopy. This is why we also aim at a setup allowing for discrete lines in the spectrum of a topological Josephson junction. The junction hybridizes the two Majorana states to form two levels differing by the parity of electron number. Therefore, the particle number-preserving interaction with microwaves does not cause transitions within this doublet. That prompts us to consider junctions of length $L\gtrsim\xi$ allowing for higher-energy Andreev states, along with the Majorana doublet (here $\xi$ is the coherence length in the topological superconductor). 

\begin{figure} \includegraphics[trim={0cm 0cm 0cm 0cm},clip,width=0.95\columnwidth]{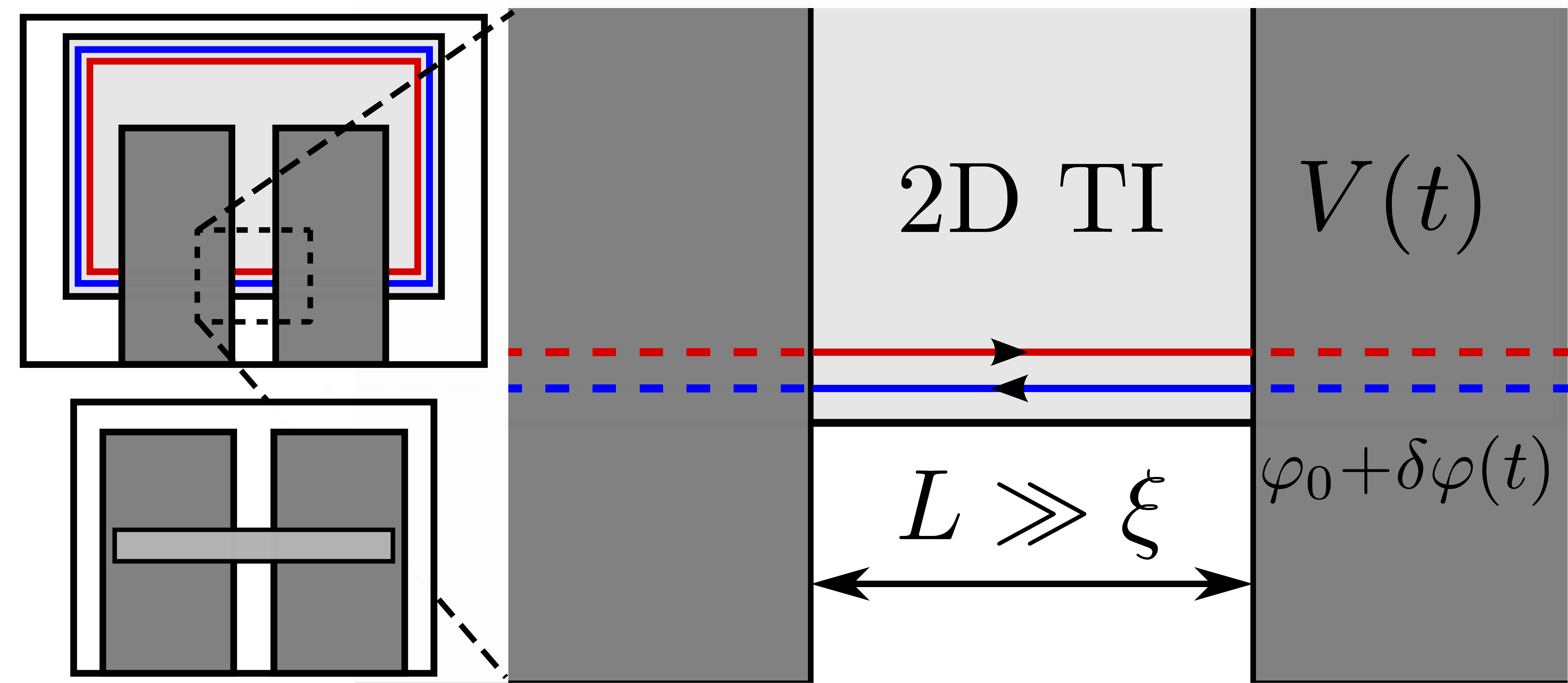} \caption{ (Color online) Right: A typical junction setup hosting Majorana  states. The arrowed lines mark the gapless helical edge modes of the 2D topological insulator (2D TI, light grey area), proximity coupled (dashed lines) to the left and right superconducting leads (grey area). The junction and the leads are wider than the coherence length $\xi$ of superconductivity induced in the TI; $\xi$ sets the scale for the bound states' decay length into the leads. A weak time-dependent voltage $V(t)$ applied between the leads gives rise to a small  component, $\delta \varphi(t)\ll 1$, modulating the superconducting phase difference across the junction (in the figure, the left lead is grounded). The resulting current response contains signatures of Majorana bound states. In the full setup (upper left inset) the outer junction can be ignored since it is much longer then the inner one and gives a signal weaker by a factor $(L/L_{\text{outer}})^2$, see Eq.~(\ref{eq:h1}). Lower inset: the Majorana detection scheme is also applicable to a junction formed by a topological semiconductor nanowire (light grey).} \label{fig:setup} \end{figure}

\begin{figure} \includegraphics[trim={0cm 0cm 0cm 0cm},clip,width=\columnwidth]{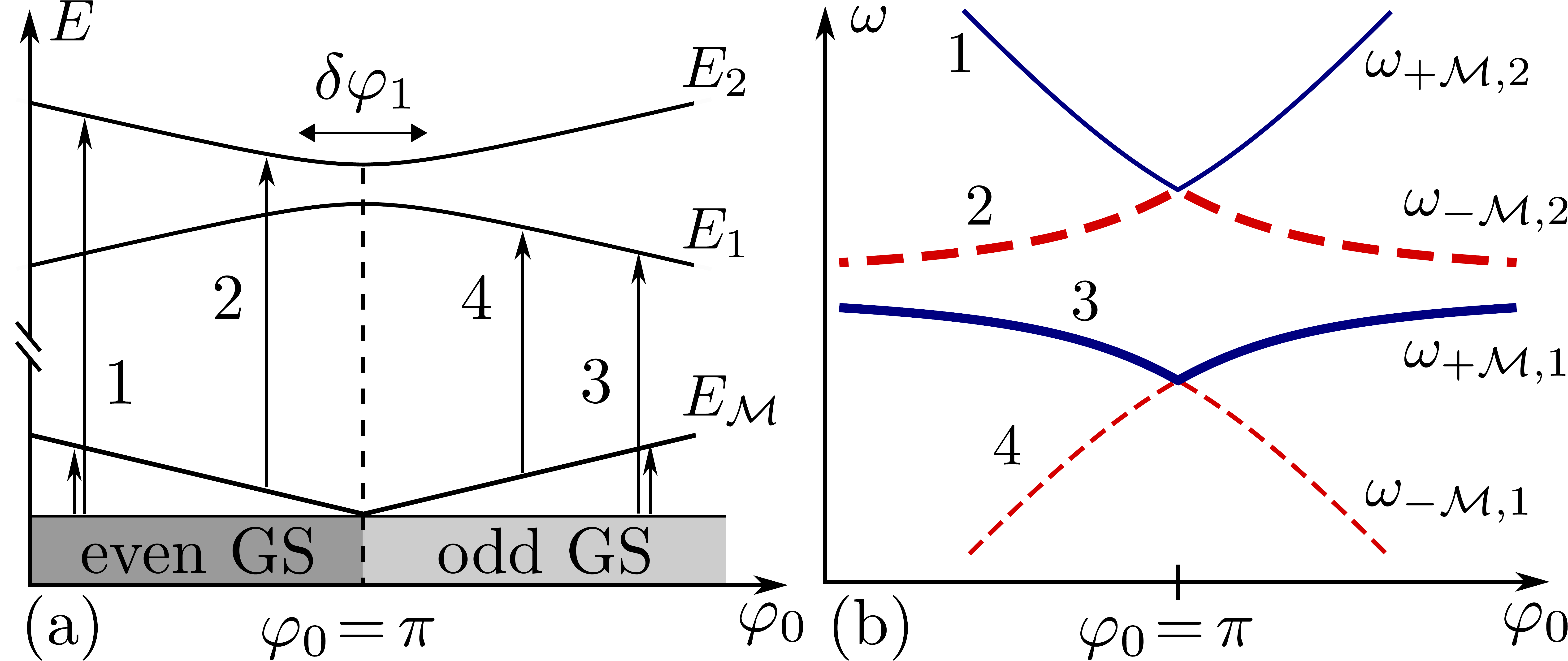} \caption{ (Color online) The low-lying excitations spectrum (a) and the absorption spectrum (b) of a topological Josephson junction. (a) The dependence of the first three Andreev levels in the junction on the phase $\varphi_0$. The lowest level has zero energy $E_\mathcal{M}$ at $\varphi_0 = \pi$, corresponding to two decoupled Majorana bound states. At $\varphi_0=\pi$ the electron number parity of the ground state (GS) changes. Energy $E_\mathcal{M}$ has a discontinuous $\varphi_0$-derivative (a kink) at the GS switching point. 
The lines $E_{1,2}(\varphi_0)$ for higher levels are smooth, as the degeneracies are lifted by disorder ($\delta\varphi_{1}$ characterizes the avoided crossing).
The kink in the $E_\mathcal{M}(\varphi_0)$ function can be probed by microwave spectroscopy. The transitions conserving the electron number parity are shown by arrows 1 through 4 and give rise to spectroscopic lines. 
(Transitions 1 and 3 create a \textit{pair} of quasiparticles above the GS.)
(b) 
 The absorption lines near $\varphi_0=\pi$, see Eq.~(\ref{omega}). At $T=0$, transitions [arrows 1 and 3 in panel (a)] start from the GS resulting in lines 1 and 3 (solid blue) each displaying a kink at $\varphi_0=\pi$.
 Populated level $E_\mathcal{M}$ enables transitions 2 and 4 (dashed red), which also show a kink, c.f. Eq.~(\ref{ReY}). 
 At finite temperature this results in a crossing of spectral lines at $\varphi_0 = \pi$.
 For the $\varphi_0$ dependence of the transitions' oscillator strengths, see Eq.~(\ref{eq:h1}).
  } \label{fig:spectrum} \end{figure}

We focus on the contribution of the discrete, sub-gap states to the admittance $Y(\omega,\varphi_0, T)$ of a topological Josephson junction~\footnote{We consider a system where the phase $\varphi_0$ is classical. Our method does not require its quantum fluctuations, unlike, for example, the proposal in Ref.~\cite{Ginossar14}.}. The setup for the junction is sketched in Fig.~\ref{fig:setup} and is based on a two-dimensional topological insulator~\cite{fu_josephson_2009} or a semiconductor nanowire~\cite{lutchyn_majorana_2010,oreg_helical_2010-1} with strong spin-orbit (SO) interaction in proximity with a conventional superconductor. The junction is controlled by the static order parameter phase difference $\varphi_0$ between the leads and is probed by applying a small voltage $V(t)$ induced by microwaves. %~\footnote{ The specific relation between $V(t)$ and the electric field amplitude of the microwaves depends on a concrete device geometry~\cite{Wang2014}}.
It creates a weak time-dependent perturbation $\delta\varphi(t)$ of the phase difference, $d(\delta\varphi)/dt =(2e/\hbar)V(t)$, which drives the transitions between the sub-gap levels. The lowest doublet is formed by the hybridized Majorana states. Their crossing at $\varphi_0=\pi$ is protected by conservation of electron number parity. (At the crossing, the number parity of the ground state switches between even and odd, see Fig.~\ref{fig:spectrum}a.) The crossings of the higher-energy Andreev levels are not protected by the parity. In a generic junction, these crossings are avoided, resulting in a smooth dependence of the corresponding energies $E_{n}(\varphi_0)$ on $\varphi_0$, see Fig.~\ref{fig:spectrum}a.

We find the discrete lines in the microwave absorption spectrum originating from the transitions in the sub-gap energy domain. The transitions involving the states of the Majorana doublet, see Fig. 2a, result in a series of lines with a characteristic kink at $\varphi_0=\pi$ in the dependence of the transition frequencies on $\varphi_0$,
\begin{equation}
\hbar\omega_{\mp \mathcal{M},n}(\varphi_0)=E_n(\pi)\mp E_\mathcal{M}^\prime |\varphi_0-\pi| +{\cal O}[(\varphi_0-\pi)^2]\,.
\label{omega}
\end{equation}
Here $\mp E_\mathcal{M}^\prime=\mp dE_\mathcal{M}/d\varphi_0$ are the slopes at $\varphi_0 = \pi$ of the energies of the two Andreev levels formed by the occupied and unoccupied Majorana doublet at the level crossing point, $E_n(\pi)$ is the energy of the higher Andreev level, and the last term in Eq.~(\ref{omega}) represents the smooth part of its $\varphi_0$-dependence. The intensity of these lines is quantified by ${\rm Re}Y(\omega,\varphi_0, T)$, for which in the vicinity of $\varphi_0=\pi$ we find the following estimate:
\begin{align}
&{\rm Re}\,Y (\omega,\varphi_0, T)
 \approx
\frac{2 \pi e^2}{\hbar}\!  \frac{(E_\mathcal{M}^\prime)^2}{\hbar\omega} \!\! \sum_{E_n>E_\mathcal{M}} \sum_{\sigma=\pm} 
\nonumber\\
&\!\times\!\!  \left[\tanh\frac{E_n(\pi)}{2T} \! - \!\tanh\frac{E_n(\pi)-\hbar\omega}{2T}\right]\!  \delta(\hbar\omega_{\sigma  \mathcal{M},n}(\varphi_0)-\hbar\omega)  .
\label{ReY}
\end{align}
Note that unlike the zero-bias anomalies in dc transport, the spectroscopic feature associated with the kink in the $\varphi_0$-dependence of $\omega_{\mp \mathcal{M},n}(\varphi_0)$  is not broadened by temperature. The brightness of lines depends weakly on temperature, with the scale provided by the higher-energy levels $E_n$. 
The admittance we find is a linear-response property of the junction; unlike the Shapiro-steps manifestation of the Majoranas, their effect on $Y$ does not set any stringent requirement on the relaxation time of the system to its equilibrium state. Our main results, Eqs.~(\ref{omega}) and (\ref{ReY}), are illustrated in Fig.~\ref{fig:spectrum}b. In the following, we outline their derivation and application to concrete junction models.

A microwave field induces voltage bias $V(t)$ between the leads of a device built on a basis of a two-dimensional topological insulator (TI) or a nanowire (NW), see Fig.~\ref{fig:setup}. The bias excites current $\left\langle I(t)\right\rangle$ between the leads, connected by the NW or the edge states of a TI. The admittance $Y(\omega)$ of the device is defined as a response function, $\left\langle I(\omega)\right\rangle =Y(\omega)V(\omega)$, at frequency $\omega$. For the current operator, we may take the current through the junction between the TI (or NW) and the right lead, $I(t)=e \cdot dN_R/dt$. Here $N_R$ is the number of electrons of the edge state in TI (or of NW) which tunneled into the right superconducting lead. 

We are interested in transitions between the states with energies below the proximity-induced gap $\Delta_0$; the latter inevitably is smaller than the superconducting gap in the leads which are the sources of proximity. That allows us to use the effective Hamiltonian of a proximized TI (or NW) instead of the full Hamiltonian in the evaluation of $dN_R/dt$~\footnote{See Supplemental Material for details. The supplement includes references to~\onlinecite{Wang2014,potter_engineering_2011,beenakker_fermion-parity_2013,zhang_josephson_2013}.}. The effective Hamiltonian $H^{(0)}=\int dx\Psi^{\dagger}\mathcal{H}^{(0)}\Psi/2$ in Nambu representation (parametrized by matrices $\tau$) takes the form 
\begin{align}
\mathcal{H}^{(0)}(x) & =-i\hbar v\tau_{z}\sigma_{z}\partial_{x}-\mu\tau_{z}+V(x)\tau_{z}+M(x)\sigma_{x}\nonumber \\
 & +\Delta_{0}(x)[\tau_{x}\cos\varphi(x)-\tau_{y}\sin\varphi(x)]\,.\label{eq:effH}
\end{align}
The first term here is the electron kinetic energy, $\mu$ is the chemical potential, $V(x)$ is the scalar potential (induced, {\it e.g.}, by disorder), and $M(x)$ is the Zeeman splitting in the junction; the $4\times 4$ matrix Hamiltonian $\mathcal{H}^{(0)}$ describes here a helical edge state in a TI~\cite{fu_josephson_2009}. (We show below that the effective Hamiltonian of a NW in a sufficiently large magnetic field takes the same form, see the paragraph preceding  Eq.~(\ref{eq:deltaphiNW}).) 
We assume the leads to be wide compared to the proximity-induced coherence length $\xi = \hbar v/\Delta_0$~\footnote{In the case of a NW junction, we assume that the ends of the wire are much farther than $\xi$ away from the junction. In the TI setup there are two junctions, see Fig.~\ref{fig:setup}. The outer one can be ignored if it is much longer than the inner one, see remark below Eq.~(\ref{eq:h1})}. 
That allows us to replace the leads depicted in Fig.~\ref{fig:setup} by semi-infinite pads,
\begin{equation}
\Delta_{0}(x)=\Delta_{0}[\Theta(-x)+\Theta(x-L)]\,,\,\,\varphi(x)=\varphi_{0}\Theta(x-L)\,,\label{eq:Gaps}
\end{equation}
where $L$ is the length of the junction. We set the order parameter of the left lead to be real so that $\varphi_{0}$ is the phase difference across the junction. Using Eq.~(\ref{eq:effH}) to evaluate $dN_R/dt$, we find
\begin{equation}
I=(2e/\hbar)(\partial H^{(0)}/\partial\varphi_0)\,.
\label{eq:current-op}
\end{equation}

Application of a bias $V(t)$ to the junction results in a time-dependent addition $\delta\varphi(t)$ to the static phase bias $\varphi_0$. According to the Josephson relation, $\delta\dot{\varphi}(t)=2eV(t)/\hbar$. For a monochromatic bias voltage $V(t)=V(\omega)\cos\omega t$, we
can treat $\delta\varphi(t)$ perturbatively as long as $|eV(\omega)/\hbar\omega|\ll1$.
(Hereinafter, we set $\hbar=1$.) In this
weak bias limit, the Hamiltonian $H$ can be split into a time-independent
part $H^{(0)}$, and a time-dependent perturbation 
\begin{equation}
H^{(1)}(t)=\delta\varphi(t)(\partial H^{(0)}/\partial\varphi_0)\,.\label{eq:pert-H}
\end{equation}
To evaluate the admittance $Y(\omega)$, we apply the standard linear response theory to the problem set by Eqs.~(\ref{eq:effH})-(\ref{eq:pert-H}). The result may be expressed in terms of the spectrum $E_n$ of  $\mathcal{H}^{(0)}$ and the matrix elements 
\begin{equation}
\mathcal{H}_{m;n}^{(1)}=\int dx\Phi_{m}^{*}(x)[\partial_{\varphi_{0}}\mathcal{H}^{(0)}(x)]\Phi_{n}(x)\label{eq:pertMatrixElem}
\end{equation}
 of the operator defining the perturbation, see Eq.~(\ref{eq:pert-H}). Here $\Phi_n (x)$ are the eigenfunctions of ${\cal H}^{(0)}$, and energies $E_n$ are
measured from the Fermi level. As the result of using the Nambu spinor notation, the energy spectrum is particle-hole symmetric (PHS), meaning that the eigenvalues come in pairs $(E_{n},\,-E_{n})$. We will label by $-n$ the state with energy $-E_{n}$ for all $n \geq 0$; the lowest ($n=0$) doublet is $(E_{\mathcal{M}},\,-E_{\mathcal{M}})$. 
 %By PHS, the matrix elements of perturbation satisfy $\mathcal{H}_{-n;m}^{(1)}=-\mathcal{H}_{-m;n}^{(1)}$. %Using the Kubo formula~\cite{LLVol5} and PHS, we find for the admittance  ($\omega^+=\omega+0 i$),  
%\begin{align}
% & Y(\omega)=\frac{i}{\omega^{+}L_{J}}+\frac{8ie^{2}}{\omega^{+}}\nonumber \\
% & \times \!\! \sum_{E_{n}>E_{m}\geq0} \left\{\frac{|\mathcal{H}_{n;m}^{(1)}|^{2}(E_{n}-E_{m})}{(\omega^{+})^2-(E_{n}-E_{m})^2}[f(E_{m})-f(E_{n})]\right.\nonumber \\
% & +\left. \frac{|\mathcal{H}_{n;-m}^{(1)}|^{2}(E_{n}+E_{m})}{(\omega^{+})^2-(E_{n}+E_{m})^2}[1-f(E_{m})-f(E_{n})]\right\}\,.\label{eq:Y-general}
%\end{align}
%Here $f(E)=1/(e^{E/T}+1)$ is the Fermi function. 
 %The first term in Eq.~(\ref{eq:Y-general}) represents the conventional inductive response of a Josephson junction. The inverse inductance $1/L_J=2e\partial_{\varphi_0}I_J$ is a derivative of the Josephson current, $I_J(\varphi_0)=\langle I\rangle_0$ with the current operator defined in Eq.~(\ref{eq:current-op}) and $\langle\dots\rangle_0$ denoting  the averaging over an equilibrium density matrix with the Hamiltonian (\ref{eq:effH}).
We concentrate on the absorption lines which originate from transitions involving a state of this doublet. Using the Kubo formula~\cite{LLVol5} and PHS, we obtain for the corresponding part of the admittance:
\begin{align}
 & {\rm Re}\,Y(\omega)=\frac{4\pi e^{2}}{\omega}\sum_{E_{n}>E_{\mathcal{M}}} \left[\tanh \frac{E_{n}}{2T}- \tanh \frac{\omega-E_{n}}{2T}\right] \nonumber  \\
 &
  \! \times\left\{|\mathcal{H}_{\mathcal{M};n}^{(1)}|^{2}\delta(\omega-\omega_{-\mathcal{M},n})+|\mathcal{H}_{-\mathcal{M};n}^{(1)}|^{2}\delta(\hbar\omega-\omega_{+\mathcal{M},n}) \right\} .\label{eq:Y-approx}
  \end{align}
Here $\omega_{\mp \mathcal{M},n} = E_{n}\mp E_{\mathcal{M}}$, see Eq.~(\ref{omega}). 
%The transitions in~(\ref{eq:Y-approx}) conserve the fermion (and quasiparticle) number parity: the first term in the brackets  arises from transitions between two different quasiparticle states, while the second term accounts for creation/annihilation of a pair of quasiparticles. There are no terms that create an odd number of quasiparticles. 
Note that there are no terms with $E_{n}=E_{\mathcal{M}}$ since  $\mathcal{H}_{\mathcal{M};-\mathcal{M}}^{(1)}=0$ by Pauli exclusion principle. 
The eigenvalues $E_{n}$ with $n \neq 0$ are in general not degenerate, contrary to the case of a conventional  time-reversal symmetric (TRS) S-N-S junction, where the states are doubly degenerate (Kramers doublets). 
The ``degeneracy'' and zero value of the energy $E_\mathcal{M}$ at $\varphi_{0}=\pi$ are protected by symmetries, and will be discussed below. % as we describe the sub-gap excitation spectrum in more detail.
 %The levels' energies and transition matrix elements in Eq.~(\ref{eq:Y-approx}) can be evaluated with the help of Eqs.~(\ref{eq:effH}), (\ref{eq:Gaps}) and (\ref{eq:pert-H}), as we show below. 

The junction sub-gap excitation spectrum is found from Eqs.~(\ref{eq:effH}) and (\ref{eq:Gaps}) by using the standard scattering matrix method~\cite{beenakker_universal_1991}.
%\begin{equation}
%\det[1-S_{N}(E)S_{A}(E)]=0\,,\label{eq:det(1-SNSA)}
%\end{equation}
%where $S_{N}$ and $S_{A}$ are respectively
%the normal and Andreev scattering matrices (defined in Appendix \ref{secAppendixDerivations}).
 We find that there is always a state with vanishing energy $E_\mathcal{M}$ at $\varphi_0 = \pi$~\cite{Note2}. The $E_\mathcal{M} =0$ state is protected by fermion number parity conservation. 
 (We assume the junction is well separated from other junctions or interfaces with bound states, allowing us to ignore hybridization of those states with $E_\mathcal{M}$, see Fig.~\ref{fig:setup} and footnote~\cite{Note3}.) 
  Near $\varphi_0 = \pi$ the dispersion $E_\mathcal{M} (\varphi_0)$ is linear. For a reflectionless junction [$M=0$ in Eq.~(\ref{eq:effH})] we find 
\begin{equation}
E_\mathcal{M} (\varphi_0)=\frac{1}{2}\frac{v}{L+\xi} |\varphi_0-\pi|\,.\label{eq:dEoverdphiBA}
\end{equation}
Breaking of TRS in the junction, ${M \neq 0}$, results in backscattering. %amplitudes $r_{+/-}(E)$ for electrons/holes at energy $E$. 
This does not lead to qualitative changes to Eq.~(\ref{eq:dEoverdphiBA}) but merely modifies the prefactor in it~\footnote{The slope is $dE_\mathcal{M}/d\varphi_0 = \pm \frac{1}{2}\frac{v}{L+\xi}(1+ {\cal O}(|r_{+}|^2)) $, see Ref.~\onlinecite{Note2}. }.

%  by a small correction of order  $|r_{\pm}(0)|^2 \ll 1$ (the slope $dE_\mathcal{M} / d \varphi_0$ for any $|r_{\pm}(0)|$ is given in Ref.~\cite{Note2}). 

The situation is different for the higher Andreev levels. 
In a reflectionless junction they have degeneracies at $\varphi_0 = \pi$: in our notation, for {\it odd} $n$ the levels $n$ and $n+1$ are degenerate, $E_{n}^{(0)}(\pi)=E_{n+1}^{(0)}(\pi)=(n+1)\pi v/2L$. 
(We take now $L \gg \xi $.) 
Backscattering lifts these degeneracies and thus leads to qualitative changes in the spectrum at $\varphi_0 =\pi$ (see Fig.~\ref{fig:spectrum}a). 
The resulting avoided crossing makes the $\varphi_0$-dependence of the levels' energies smooth~\footnote{Equations~(\ref{eq:dEoverdphiBA}) and (\ref{eq:En}) are valid in the full range of $\varphi_0$ except for a narrow  interval around $\varphi_0 =0$ where a similar avoided crossing happens.} 
\begin{equation}
E_{n} (\varphi_0)=E_{n}^{(0)}(\pi) + (-1)^n \frac{v}{2L} \sqrt{(\varphi_0-\pi)^2+(\delta\varphi_n)^2}\,.\label{eq:En}
\end{equation}
(We ignore here corrections to the prefactor $v/2L$ due to weak  backscattering.) 
%\begin{equation}
%E_{n}(\varphi_{0})=\begin{cases}
%\frac{v}{L}n\pi-\frac{v}{2L}\sqrt{(\varphi_{0}-\pi)^{2}+(\delta\varphi_{n})^{2}}\,, & n\text{ odd},\\
%\frac{v}{L}(n-1)\pi+\frac{v}{2L}\sqrt{(\varphi_{0}-\pi)^{2}+(\delta\varphi_{n-1})^{2}}\,, & n\text{ even},
%\end{cases}
%\end{equation}
The smoothness in~(\ref{eq:En}) is characterized by the width of the avoided-crossing region $\delta\varphi_n= \left|r_{+}+r_{-}^{*}\right|_{E=E_{n}^{(0)}(\pi)}$. We denote by $r_{+/-}^{(\prime)}(E)$ the reflection amplitudes for electrons/holes entering the junction from the left (right) at energy $E$.  %, determined by the reflection amplitudes at energy $E=E_{n}^{(0)}(\pi)$.
 In a simple model with ${V(x)=0}$,  $M(x)=M\Theta(x)\Theta{(L-x)}$, and $M \ll v/L$, the junction is symmetric  and  %$r_{\pm}(E)=r_{\pm}^{\prime}(E)=-i\frac{M}{v/L}\frac{\pm\mu}{E\pm\mu}$,  valid for energies which are integer multiples of $\pi v /L$. 
%$r_{\pm} \! =\!r^{\prime}_{\pm} \!=\!-M(e^{2i(E\pm\mu)L/v}\!-\!1)/2(E\!\pm\!\mu)$. 
$r_{\pm}(E) \!\!=\!M{(1\!-\!e^{2i(E\pm\mu)L/v})}/{2(E\pm\mu)}$.  
In this model,  for small chemical potential,  $\mu \ll v/L$, the width of the avoided crossing is~\footnote{Lifting of degeneracies in the  model for which Eq.~(\ref{eq:deltaphiSimple}) was derived requires, in addition to $M\neq 0$, a non-zero chemical potential $\mu$, due to special symmetries~\cite{Note2} of~(\ref{eq:effH}) at the Dirac point.}  %~\footnote{Lifting of degeneracies in the junction model for which Eq.~(\ref{eq:deltaphiSimple}) was derived requires, in addition to $M$, a non-zero chemical potential. \textbf{This is because of special symmetries~\cite{Note2} of~(\ref{eq:effH}) under particle-hole transformation and reflection about junction center}.  These symmetries protect the degeneracies at $\varphi_0 = \pi$ at the Dirac point.} 
\begin{equation}
\delta\varphi_{n}=\frac{|M|}{v/2L}\frac{|\mu|}{E_{n}^{(0)}(\pi)}\,.\label{eq:deltaphiSimple}
\end{equation}
  
The sub-gap spectrum determines the resonance frequencies $\omega_{\mp \mathcal{M},n} (\varphi_0)$: combining Eqs.~(\ref{eq:dEoverdphiBA}) and (\ref{eq:En}) yields Eq.~(\ref{omega}).
The brightness of the spectral lines is set by the matrix elements in Eq.~(\ref{eq:Y-approx}). 
They %transition matrix elements $\mathcal{H}_{\pm \mathcal{M};n}^{(1)}$ 
are calculated from Eq.~(\ref{eq:pertMatrixElem}) where, according to Eq.~(\ref{eq:Gaps}), the integration over $x$ is restricted to the right superconducting region $x>L$ where $\partial_{\varphi_{0}}\mathcal{H}^{(0)}(x)$ is non-zero. There, the sub-gap wave functions decay exponentially, $\Phi_n (x)=e^{-(x-L)/\xi}\Phi_n (L)$. We assume here $E_n \ll \Delta_0$ so that the decay length is approximated by $\xi$. 
In the limit of weak reflection, we find~\cite{Note2} 
\begin{equation}
|\mathcal{H}_{\pm \mathcal{M};n}^{(1)}|^{2} \! =\!\!\frac{v^{2}}{8L^{2}}\!\!\left[1\pm\! (-1)^n \frac{|\varphi_{0}-\pi|\pm 2\text{Re}\delta\Phi_{\mathcal{M}}^* \delta\varphi_{n} e^{i\vartheta_n}}{\sqrt{(\delta\varphi_{n})^{2}+(\varphi_{0}-\pi)^{2}}}\right]\!.
\label{eq:h1}
\end{equation}
Importantly, $|\mathcal{H}_{\pm \mathcal{M};n}^{(1)}|^{2} \propto 1/L^2$; likewise, the contribution in the admittance from the long outer junction in Fig.~\ref{fig:setup} is proportional to $ 1/L_\text{outer}^2$ and can be  ignored since the corresponding length $L_\text{outer} \gg L$. 
In Eq.~(\ref{eq:h1}) $\vartheta_n=\arg(r_{+}^{\prime}+r_{-}^{\prime*})_{E=E_{n}^{(0)}(\pi)}$ and  $\delta\Phi_{\mathcal{M}}=\frac{v}{2L}[d_{+}^{2}\partial_{E}r_{+}^{*}]_{E=0}$ is the small correction (due to backscattering) to the wave function of level $E_\mathcal{M}$; here $d_{\pm}$ is the transmission amplitude.  
In the simple model used to derive Eq.~(\ref{eq:deltaphiSimple}),  $\delta\Phi_{\mathcal{M}}=\frac{M}{v/L}{(\frac{1}{2}+\frac{i}{3}\frac{\mu}{v/L})}$.
% $\text{Re}\delta\Phi_{\mathcal{M}}^{*}e^{i\vartheta_n}=\frac{1}{3} \frac{|M||\mu|}{(v/L)^2}$

Since Eq.~(\ref{eq:h1}) was derived assuming $L \gg \xi$, we can express its prefactor as $(E_\mathcal{M}^{\prime})^2/2$ by using Eq.~(\ref{eq:dEoverdphiBA}) in that limit; the replacement is valid if the levels resolved in the transitions are far below $\Delta_0$. 
%Furthermore, the second term in the square brackets in~(\ref{eq:h1}) is negligible since $\delta \Phi_\mathcal{M} \ll 1$ and  $|\varphi_0 - \pi| \ll \delta \varphi_n$. 
Furthermore, close to $\varphi_0 = \pi$ we can neglect the second term in the square brackets in~(\ref{eq:h1}) as long as $|\varphi_0 - \pi| \ll \delta \varphi_n$, since $|\delta \Phi_\mathcal{M}| \ll 1$. 
Using the approximate matrix elements in Eq.~(\ref{eq:Y-approx}) leads to Eq.~(\ref{ReY}). 

Approximating the matrix elements in Eq.~(\ref{ReY}) by a constant $(E_\mathcal{M}^{\prime})^2/2$ does not capture the brightness variations of the absorption lines with $n$ and $\varphi_0$. 
 At $\varphi_0 = \pi$ 
% the lines 3 and 4 (transitions to level $E_1$)  in Fig.~\ref{fig:spectrum}b are brighter than lines 1 and 2 (transitions to level $E_2$).
 transitions to levels $E_{n}$ with \textit{odd} $n$ (lines 3 and 4 in  Fig.~\ref{fig:spectrum}b) are brighter than transitions to levels $E_{n+1}$ (lines 1 and 2), due to $\delta \Phi_\mathcal{M} \neq 0$.
 As a function of $\varphi_0$, the brightness is non-analytic at $\varphi_0 = \pi$ which is a maximum (minimum) for lines 1 and 4 (2 and 3) in Fig.~\ref{fig:spectrum}b. 
 Far from the avoided crossing, $|\varphi_0 - \pi|\gg \delta \varphi_n$, the spectral lines 1 and 4 become dim. In general, the lines corresponding to frequencies $\omega_{-\mathcal{M},n}$ and $\omega_{+\mathcal{M},n+1}$ with $n$ {\it odd} become dim away from $\varphi_0 = \pi$.
 This is because of an approximate selection rule for the matrix elements $\mathcal{H}_{\pm \mathcal{M};n}^{(1)}$: 
  we see from Eq.~(\ref{eq:h1}) that $|\mathcal{H}_{\mathcal{M} ;n}^{(1)}|^2$ and $|\mathcal{H}_{-\mathcal{M};n+1}^{(1)}|^2$ for odd $n$ are smaller by factor
  $\propto \! \text{Re}\delta\Phi_{\mathcal{M}}^* \delta\varphi_{n} e^{i\vartheta_n}/ |\varphi_0 - \pi|$
 %$\propto \! \delta \varphi_n \delta \Phi_\mathcal{M} / |\varphi_0 - \pi|$ 
   compared to those with even $n$.~\footnote{This selection rule is exact in a reflectionless junction, $\delta \varphi_n \!=\! 0$, where the operator $S_z$ commutes with ${H^{(0)}}$ and $H^{(1)}$. %Because of the spin-momentum locking in Eq.~(\ref{eq:effH}), the selection rule is seen as a ``conservation of slope'' in Fig.~\ref{fig:spectrum}a.}
}
 This analysis allows us to extrapolate our theory to stronger backscattering, $\delta \varphi_n \sim 1$. 
 Due to the large width of the avoided crossing, the above approximate selection rule becomes inapplicable. 
 On the other hand, at $\varphi_0 = \pi$  the alternation of the lines' intensities with $n$ becomes more pronounced. 
   The main feature, the kink in the transition frequencies $\omega_{\mp \mathcal{M},n}(\varphi_0)$ at $\varphi_0 = \pi$, persists. 
 %, but becomes flatter due to decreasing slope of $E_\mathcal{M}$.

 %The behavior of a junction where the backscattering probability is not small can be extrapolated from the above considerations. The main feature, the kink, is preserved. The width of the avoided crossing region becomes larger, $\delta \varphi_n \sim 1$. Likewise, the contrast 

%NW junction

We derived Eqs.~(\ref{omega}) and (\ref{ReY}) for a TI junction, but the same low-energy model, Eqs.~(\ref{eq:effH})-(\ref{eq:En}), is applicable to a NW-based setup. For illustration, we concentrate here on the limit of large Zeeman energy, $B\gg m\alpha^2$; the SO energy scale here is
determined by the electron effective mass $m$ and SO velocity $\alpha$. At low-energies $E\ll B$, linearization~\cite{klinovaja_composite_2012} of the spectrum near the Fermi points $k\!\approx\!\pm k_{Z}\!=\!\pm \sqrt{2mB}$ leads to an effective Hamiltonian of the form~(\ref{eq:effH}) with $v\!=\!k_{Z}/m$, and $\Delta_{0}\!=\!\Delta 2m\alpha/k_Z$ where $\Delta$ is the induced s-wave gap in the nanowire. (We set $\mu=0$ for simplicity.)
The structure of spectrum of a long NW junction is therefore identical to that of the TI junction described above -- only the microscopic forms of the phenomenological parameters in~(\ref{eq:effH}) are different.

%In the NW junction non-zero reflection amplitudes, and lifting of degeneracies of higher Andreev levels, may originate even from time-reversal symmetric scalar disorder.
In the NW junction even TRS scalar disorder may lift the degeneracies of high Andreev levels. 
We illustrate this by considering a short-range impurity $u_0 \delta(x-L)$ in the microscopic Hamiltonian. (For definiteness, we take the impurity to be at the junction interface, as was done in Ref.~\onlinecite{lutchyn_majorana_2010}.) 
The low-energy projection of the impurity Hamiltonian yields respective forward and backscattering terms~\cite{Note2} $V(x) \approx u_{0} \delta(x-L)$ and $M(x)\approx u_{0} \delta(x-L)$ in Eq.~(\ref{eq:effH}). 
(The prefactors here are given to lowest order in $m \alpha^2/B$.) 
%\begin{equation}
%\mathcal{H}_{\text{imp}}=l\delta(x-L)[V_{0}\tau_{z}+M_{0}\sigma_{x}]\,,
%\label{eq:NWimpurity}
%\end{equation}
%where the first and second terms correspond to respective forward and backscattering~\cite{Note2}. 
The width of the avoided crossing,  
\begin{equation}
\delta\varphi_{n}=2\frac{|u_0|}{v}\frac{\xi}{L}\frac{ E_{n}^{(0)}(\pi)}{v/L}\,,\label{eq:deltaphiNW}
\end{equation}
is finite due to $M\neq 0$; here, as before, for $n$ odd  $E_{n}^{(0)}(\pi)\!\!=\!\!E_{n+1}^{(0)}(\pi)\!\!=\!\!{(n+1)}\pi v/2(L+\xi)$. 
 The correction to the lowest-level wave function is  %$\delta\Phi_{\mathcal{M}}=()\frac{u_0}{v}(1-i\frac{u_0}{v})$. 
 $\delta\Phi_{\mathcal{M}}=(u_0/v)(1-iu_0/v)$ and the phase appearing  in Eq.~(\ref{eq:h1}) is
 $e^{i\vartheta_n} = (u_0/|u_0|)(1-iu_0/v)$. 
We see that the avoided crossings and the transition matrix elements in a NW junction can be quantified in the same way as in a TI junction.

Using Eq.~(\ref{eq:dEoverdphiBA}), we may compare the strength of absorption, Eq.~(\ref{ReY}), with the corresponding strength~\cite{Kos13} of transition within the pair of Andreev levels in a conventional short S-N-S junction~\cite{bretheau2013a,Bretheau2013b}. At equal frequencies $\omega$, the transition in the ``Majorana'' junction is stronger than the one in a conventional junction if the transmission coefficient of the latter is $<0.4$.

In summary, we have shown how Majorana bound states manifest in the finite-frequency admittance of a topological Josephson junction with multiple Andreev levels. 
Our main finding is a kink in the $\varphi_0$-dependence of the resonant absorption frequency, and can be observed in the dissipative (real) part of the admittance $Y(\omega)$, see Eq.~(\ref{omega}) and Fig.~\ref{fig:spectrum}b. 
Alternatively, one may employ the reflected microwaves' phase shift, $\arg Y(\omega)$, obtained from  Eq.~(\ref{eq:Y-approx}) and its Kramers-Kronig partner. 
The frequency-dependent phase shift jumps by $\pi$ across a resonance, and the position of this jump as a function of $\varphi_0$ shows a kink. 
The kink is a consequence of the ground state parity switching in the junction, or decoupling of Majorana states. 
The admittance provides a novel, linear-response signature of this decoupling.  

\begin{acknowledgments}
We thank Michel Devoret and Liang Fu for discussions, and Richard Brierley and Hendrik Meier for valuable comments on the manuscript. 
This work was supported by NSF DMR Grant 1206612, 
ONR Grant Q00704, ARO Grant W911NF-09-1-0514, DFG through SFB 767, EU FP7 Marie Curie Zukunftskolleg Incoming Fellowship Programme  (Grant 291784) and the Senior Fellowship of the Zukunftskolleg, Konstanz. 
\end{acknowledgments}

\bibliographystyle{apsrev4-1}
\bibliography{STIS_Refs}

\begin{thebibliography}{40}%
\makeatletter
\providecommand \@ifxundefined [1]{%
 \@ifx{#1\undefined}
}%
\providecommand \@ifnum [1]{%
 \ifnum #1\expandafter \@firstoftwo
 \else \expandafter \@secondoftwo
 \fi
}%
\providecommand \@ifx [1]{%
 \ifx #1\expandafter \@firstoftwo
 \else \expandafter \@secondoftwo
 \fi
}%
\providecommand \natexlab [1]{#1}%
\providecommand \enquote  [1]{``#1''}%
\providecommand \bibnamefont  [1]{#1}%
\providecommand \bibfnamefont [1]{#1}%
\providecommand \citenamefont [1]{#1}%
\providecommand \href@noop [0]{\@secondoftwo}%
\providecommand \href [0]{\begingroup \@sanitize@url \@href}%
\providecommand \@href[1]{\@@startlink{#1}\@@href}%
\providecommand \@@href[1]{\endgroup#1\@@endlink}%
\providecommand \@sanitize@url [0]{\catcode `\\12\catcode `\$12\catcode
  `\&12\catcode `\#12\catcode `\^12\catcode `\_12\catcode `\%12\relax}%
\providecommand \@@startlink[1]{}%
\providecommand \@@endlink[0]{}%
\providecommand \url  [0]{\begingroup\@sanitize@url \@url }%
\providecommand \@url [1]{\endgroup\@href {#1}{\urlprefix }}%
\providecommand \urlprefix  [0]{URL }%
\providecommand \Eprint [0]{\href }%
\providecommand \doibase [0]{http://dx.doi.org/}%
\providecommand \selectlanguage [0]{\@gobble}%
\providecommand \bibinfo  [0]{\@secondoftwo}%
\providecommand \bibfield  [0]{\@secondoftwo}%
\providecommand \translation [1]{[#1]}%
\providecommand \BibitemOpen [0]{}%
\providecommand \bibitemStop [0]{}%
\providecommand \bibitemNoStop [0]{.\EOS\space}%
\providecommand \EOS [0]{\spacefactor3000\relax}%
\providecommand \BibitemShut  [1]{\csname bibitem#1\endcsname}%
\let\auto@bib@innerbib\@empty
%</preamble>
\bibitem [{\citenamefont {Kitaev}(2003)}]{Kitaev20032}%
  \BibitemOpen
  \bibfield  {author} {\bibinfo {author} {\bibfnamefont {A.}~\bibnamefont
  {Kitaev}},\ }\href {\doibase http://dx.doi.org/10.1016/S0003-4916(02)00018-0}
  {\bibfield  {journal} {\bibinfo  {journal} {Annals of Physics}\ }\textbf
  {\bibinfo {volume} {303}},\ \bibinfo {pages} {2 } (\bibinfo {year}
  {2003})}\BibitemShut {NoStop}%
\bibitem [{\citenamefont {{Nayak}}\ \emph {et~al.}(2008)\citenamefont
  {{Nayak}}, \citenamefont {{Simon}}, \citenamefont {{Stern}}, \citenamefont
  {{Freedman}},\ and\ \citenamefont {{Das Sarma}}}]{2008RvMP...80.1083N}%
  \BibitemOpen
  \bibfield  {author} {\bibinfo {author} {\bibfnamefont {C.}~\bibnamefont
  {{Nayak}}}, \bibinfo {author} {\bibfnamefont {S.~H.}\ \bibnamefont
  {{Simon}}}, \bibinfo {author} {\bibfnamefont {A.}~\bibnamefont {{Stern}}},
  \bibinfo {author} {\bibfnamefont {M.}~\bibnamefont {{Freedman}}}, \ and\
  \bibinfo {author} {\bibfnamefont {S.}~\bibnamefont {{Das Sarma}}},\ }\href
  {\doibase 10.1103/RevModPhys.80.1083} {\bibfield  {journal} {\bibinfo
  {journal} {Reviews of Modern Physics}\ }\textbf {\bibinfo {volume} {80}},\
  \bibinfo {pages} {1083} (\bibinfo {year} {2008})},\ \Eprint
  {http://arxiv.org/abs/0707.1889} {arXiv:0707.1889 [cond-mat.str-el]}
  \BibitemShut {NoStop}%
\bibitem [{\citenamefont {{Alicea}}(2012)}]{2012RPPh...75g6501A}%
  \BibitemOpen
  \bibfield  {author} {\bibinfo {author} {\bibfnamefont {J.}~\bibnamefont
  {{Alicea}}},\ }\href {\doibase 10.1088/0034-4885/75/7/076501} {\bibfield
  {journal} {\bibinfo  {journal} {Reports on Progress in Physics}\ }\textbf
  {\bibinfo {volume} {75}},\ \bibinfo {eid} {076501} (\bibinfo {year}
  {2012})},\ \Eprint {http://arxiv.org/abs/1202.1293} {arXiv:1202.1293
  [cond-mat.supr-con]} \BibitemShut {NoStop}%
\bibitem [{\citenamefont {Beenakker}(2013)}]{Beenakker13}%
  \BibitemOpen
  \bibfield  {author} {\bibinfo {author} {\bibfnamefont {C.}~\bibnamefont
  {Beenakker}},\ }\href {\doibase 10.1146/annurev-conmatphys-030212-184337}
  {\bibfield  {journal} {\bibinfo  {journal} {Annual Review of Condensed Matter
  Physics}\ }\textbf {\bibinfo {volume} {4}},\ \bibinfo {pages} {113} (\bibinfo
  {year} {2013})},\ \Eprint
  {http://arxiv.org/abs/http://dx.doi.org/10.1146/annurev-conmatphys-030212-184337}
  {http://dx.doi.org/10.1146/annurev-conmatphys-030212-184337} \BibitemShut
  {NoStop}%
\bibitem [{\citenamefont {Mourik}\ \emph {et~al.}(2012)\citenamefont {Mourik},
  \citenamefont {Zuo}, \citenamefont {Frolov}, \citenamefont {Plissard},
  \citenamefont {Bakkers},\ and\ \citenamefont
  {Kouwenhoven}}]{mourik_signatures_2012}%
  \BibitemOpen
  \bibfield  {author} {\bibinfo {author} {\bibfnamefont {V.}~\bibnamefont
  {Mourik}}, \bibinfo {author} {\bibfnamefont {K.}~\bibnamefont {Zuo}},
  \bibinfo {author} {\bibfnamefont {S.~M.}\ \bibnamefont {Frolov}}, \bibinfo
  {author} {\bibfnamefont {S.~R.}\ \bibnamefont {Plissard}}, \bibinfo {author}
  {\bibfnamefont {E.~P. A.~M.}\ \bibnamefont {Bakkers}}, \ and\ \bibinfo
  {author} {\bibfnamefont {L.~P.}\ \bibnamefont {Kouwenhoven}},\ }\href
  {\doibase 10.1126/science.1222360} {\bibfield  {journal} {\bibinfo  {journal}
  {Science}\ }\textbf {\bibinfo {volume} {336}},\ \bibinfo {pages} {1003}
  (\bibinfo {year} {2012})}\BibitemShut {NoStop}%
\bibitem [{\citenamefont {Rokhinson}\ \emph {et~al.}(2012)\citenamefont
  {Rokhinson}, \citenamefont {Liu},\ and\ \citenamefont
  {Furdyna}}]{rokhinson_fractional_2012}%
  \BibitemOpen
  \bibfield  {author} {\bibinfo {author} {\bibfnamefont {L.~P.}\ \bibnamefont
  {Rokhinson}}, \bibinfo {author} {\bibfnamefont {X.}~\bibnamefont {Liu}}, \
  and\ \bibinfo {author} {\bibfnamefont {J.~K.}\ \bibnamefont {Furdyna}},\
  }\href {\doibase 10.1038/nphys2429} {\bibfield  {journal} {\bibinfo
  {journal} {Nature Physics}\ }\textbf {\bibinfo {volume} {8}},\ \bibinfo
  {pages} {795} (\bibinfo {year} {2012})}\BibitemShut {NoStop}%
\bibitem [{\citenamefont {Deng}\ \emph {et~al.}(2012)\citenamefont {Deng},
  \citenamefont {Yu}, \citenamefont {Huang}, \citenamefont {Larsson},
  \citenamefont {Caroff},\ and\ \citenamefont {Xu}}]{Deng2012}%
  \BibitemOpen
  \bibfield  {author} {\bibinfo {author} {\bibfnamefont {M.~T.}\ \bibnamefont
  {Deng}}, \bibinfo {author} {\bibfnamefont {C.~L.}\ \bibnamefont {Yu}},
  \bibinfo {author} {\bibfnamefont {G.~Y.}\ \bibnamefont {Huang}}, \bibinfo
  {author} {\bibfnamefont {M.}~\bibnamefont {Larsson}}, \bibinfo {author}
  {\bibfnamefont {P.}~\bibnamefont {Caroff}}, \ and\ \bibinfo {author}
  {\bibfnamefont {H.~Q.}\ \bibnamefont {Xu}},\ }\href {\doibase
  10.1021/nl303758w} {\bibfield  {journal} {\bibinfo  {journal} {Nano Letters}\
  }\textbf {\bibinfo {volume} {12}},\ \bibinfo {pages} {6414} (\bibinfo {year}
  {2012})},\ \bibinfo {note} {pMID: 23181691},\ \Eprint
  {http://arxiv.org/abs/http://dx.doi.org/10.1021/nl303758w}
  {http://dx.doi.org/10.1021/nl303758w} \BibitemShut {NoStop}%
\bibitem [{\citenamefont {{Das}}\ \emph {et~al.}(2012)\citenamefont {{Das}},
  \citenamefont {{Ronen}}, \citenamefont {{Most}}, \citenamefont {{Oreg}},
  \citenamefont {{Heiblum}},\ and\ \citenamefont
  {{Shtrikman}}}]{2012NatPh...8..887D}%
  \BibitemOpen
  \bibfield  {author} {\bibinfo {author} {\bibfnamefont {A.}~\bibnamefont
  {{Das}}}, \bibinfo {author} {\bibfnamefont {Y.}~\bibnamefont {{Ronen}}},
  \bibinfo {author} {\bibfnamefont {Y.}~\bibnamefont {{Most}}}, \bibinfo
  {author} {\bibfnamefont {Y.}~\bibnamefont {{Oreg}}}, \bibinfo {author}
  {\bibfnamefont {M.}~\bibnamefont {{Heiblum}}}, \ and\ \bibinfo {author}
  {\bibfnamefont {H.}~\bibnamefont {{Shtrikman}}},\ }\href {\doibase
  10.1038/nphys2479} {\bibfield  {journal} {\bibinfo  {journal} {Nature
  Physics}\ }\textbf {\bibinfo {volume} {8}},\ \bibinfo {pages} {887} (\bibinfo
  {year} {2012})},\ \Eprint {http://arxiv.org/abs/1205.7073} {arXiv:1205.7073
  [cond-mat.mes-hall]} \BibitemShut {NoStop}%
\bibitem [{\citenamefont {{Koren}}\ \emph {et~al.}(2011)\citenamefont
  {{Koren}}, \citenamefont {{Kirzhner}}, \citenamefont {{Lahoud}},
  \citenamefont {{Chashka}},\ and\ \citenamefont
  {{Kanigel}}}]{2011PhRvB..84v4521K}%
  \BibitemOpen
  \bibfield  {author} {\bibinfo {author} {\bibfnamefont {G.}~\bibnamefont
  {{Koren}}}, \bibinfo {author} {\bibfnamefont {T.}~\bibnamefont {{Kirzhner}}},
  \bibinfo {author} {\bibfnamefont {E.}~\bibnamefont {{Lahoud}}}, \bibinfo
  {author} {\bibfnamefont {K.~B.}\ \bibnamefont {{Chashka}}}, \ and\ \bibinfo
  {author} {\bibfnamefont {A.}~\bibnamefont {{Kanigel}}},\ }\href {\doibase
  10.1103/PhysRevB.84.224521} {\bibfield  {journal} {\bibinfo  {journal}
  {\prb}\ }\textbf {\bibinfo {volume} {84}},\ \bibinfo {eid} {224521} (\bibinfo
  {year} {2011})},\ \Eprint {http://arxiv.org/abs/1111.3445} {arXiv:1111.3445
  [cond-mat.supr-con]} \BibitemShut {NoStop}%
\bibitem [{\citenamefont {{Kurter}}\ \emph {et~al.}(2013)\citenamefont
  {{Kurter}}, \citenamefont {{Finck}}, \citenamefont {{Hor}},\ and\
  \citenamefont {{Van Harlingen}}}]{Kurter13}%
  \BibitemOpen
  \bibfield  {author} {\bibinfo {author} {\bibfnamefont {C.}~\bibnamefont
  {{Kurter}}}, \bibinfo {author} {\bibfnamefont {A.~D.~K.}\ \bibnamefont
  {{Finck}}}, \bibinfo {author} {\bibfnamefont {Y.~S.}\ \bibnamefont {{Hor}}},
  \ and\ \bibinfo {author} {\bibfnamefont {D.~J.}\ \bibnamefont {{Van
  Harlingen}}},\ }\href@noop {} {\bibfield  {journal} {\bibinfo  {journal}
  {ArXiv e-prints}\ } (\bibinfo {year} {2013})},\ \Eprint
  {http://arxiv.org/abs/1307.7764} {arXiv:1307.7764 [cond-mat.mes-hall]}
  \BibitemShut {NoStop}%
\bibitem [{\citenamefont {{Hart}}\ \emph {et~al.}(2014)\citenamefont {{Hart}},
  \citenamefont {{Ren}}, \citenamefont {{Wagner}}, \citenamefont {{Leubner}},
  \citenamefont {{M{\"u}hlbauer}}, \citenamefont {{Br{\"u}ne}}, \citenamefont
  {{Buhmann}}, \citenamefont {{Molenkamp}},\ and\ \citenamefont
  {{Yacoby}}}]{hart_induced_2013}%
  \BibitemOpen
  \bibfield  {author} {\bibinfo {author} {\bibfnamefont {S.}~\bibnamefont
  {{Hart}}}, \bibinfo {author} {\bibfnamefont {H.}~\bibnamefont {{Ren}}},
  \bibinfo {author} {\bibfnamefont {T.}~\bibnamefont {{Wagner}}}, \bibinfo
  {author} {\bibfnamefont {P.}~\bibnamefont {{Leubner}}}, \bibinfo {author}
  {\bibfnamefont {M.}~\bibnamefont {{M{\"u}hlbauer}}}, \bibinfo {author}
  {\bibfnamefont {C.}~\bibnamefont {{Br{\"u}ne}}}, \bibinfo {author}
  {\bibfnamefont {H.}~\bibnamefont {{Buhmann}}}, \bibinfo {author}
  {\bibfnamefont {L.~W.}\ \bibnamefont {{Molenkamp}}}, \ and\ \bibinfo {author}
  {\bibfnamefont {A.}~\bibnamefont {{Yacoby}}},\ }\href {\doibase
  10.1038/nphys3036} {\bibfield  {journal} {\bibinfo  {journal} {Nature
  Physics}\ }\textbf {\bibinfo {volume} {10}},\ \bibinfo {pages} {638}
  (\bibinfo {year} {2014})},\ \Eprint {http://arxiv.org/abs/1312.2559}
  {arXiv:1312.2559 [cond-mat.mes-hall]} \BibitemShut {NoStop}%
\bibitem [{\citenamefont {{Pribiag}}\ \emph {et~al.}(2014)\citenamefont
  {{Pribiag}}, \citenamefont {{Beukman}}, \citenamefont {{Qu}}, \citenamefont
  {{Cassidy}}, \citenamefont {{Charpentier}}, \citenamefont {{Wegscheider}},\
  and\ \citenamefont {{Kouwenhoven}}}]{Pribiag14}%
  \BibitemOpen
  \bibfield  {author} {\bibinfo {author} {\bibfnamefont {V.~S.}\ \bibnamefont
  {{Pribiag}}}, \bibinfo {author} {\bibfnamefont {A.~J.~A.}\ \bibnamefont
  {{Beukman}}}, \bibinfo {author} {\bibfnamefont {F.}~\bibnamefont {{Qu}}},
  \bibinfo {author} {\bibfnamefont {M.~C.}\ \bibnamefont {{Cassidy}}}, \bibinfo
  {author} {\bibfnamefont {C.}~\bibnamefont {{Charpentier}}}, \bibinfo {author}
  {\bibfnamefont {W.}~\bibnamefont {{Wegscheider}}}, \ and\ \bibinfo {author}
  {\bibfnamefont {L.~P.}\ \bibnamefont {{Kouwenhoven}}},\ }\href@noop {}
  {\bibfield  {journal} {\bibinfo  {journal} {ArXiv e-prints}\ } (\bibinfo
  {year} {2014})},\ \Eprint {http://arxiv.org/abs/1408.1701} {arXiv:1408.1701
  [cond-mat.mes-hall]} \BibitemShut {NoStop}%
\bibitem [{\citenamefont {Law}\ \emph {et~al.}(2009)\citenamefont {Law},
  \citenamefont {Lee},\ and\ \citenamefont {Ng}}]{law2009}%
  \BibitemOpen
  \bibfield  {author} {\bibinfo {author} {\bibfnamefont {K.~T.}\ \bibnamefont
  {Law}}, \bibinfo {author} {\bibfnamefont {P.~A.}\ \bibnamefont {Lee}}, \ and\
  \bibinfo {author} {\bibfnamefont {T.~K.}\ \bibnamefont {Ng}},\ }\href
  {\doibase 10.1103/PhysRevLett.103.237001} {\bibfield  {journal} {\bibinfo
  {journal} {Phys. Rev. Lett.}\ }\textbf {\bibinfo {volume} {103}},\ \bibinfo
  {pages} {237001} (\bibinfo {year} {2009})}\BibitemShut {NoStop}%
\bibitem [{\citenamefont {Flensberg}(2010)}]{flensberg_tunneling_2010}%
  \BibitemOpen
  \bibfield  {author} {\bibinfo {author} {\bibfnamefont {K.}~\bibnamefont
  {Flensberg}},\ }\href {\doibase 10.1103/PhysRevB.82.180516} {\bibfield
  {journal} {\bibinfo  {journal} {Phys. Rev. B}\ }\textbf {\bibinfo {volume}
  {82}},\ \bibinfo {pages} {180516} (\bibinfo {year} {2010})}\BibitemShut
  {NoStop}%
\bibitem [{\citenamefont {{Sau}}\ \emph {et~al.}(2010)\citenamefont {{Sau}},
  \citenamefont {{Tewari}}, \citenamefont {{Lutchyn}}, \citenamefont
  {{Stanescu}},\ and\ \citenamefont {{Das Sarma}}}]{Sau2010a}%
  \BibitemOpen
  \bibfield  {author} {\bibinfo {author} {\bibfnamefont {J.~D.}\ \bibnamefont
  {{Sau}}}, \bibinfo {author} {\bibfnamefont {S.}~\bibnamefont {{Tewari}}},
  \bibinfo {author} {\bibfnamefont {R.~M.}\ \bibnamefont {{Lutchyn}}}, \bibinfo
  {author} {\bibfnamefont {T.~D.}\ \bibnamefont {{Stanescu}}}, \ and\ \bibinfo
  {author} {\bibfnamefont {S.}~\bibnamefont {{Das Sarma}}},\ }\href {\doibase
  10.1103/PhysRevB.82.214509} {\bibfield  {journal} {\bibinfo  {journal}
  {\prb}\ }\textbf {\bibinfo {volume} {82}},\ \bibinfo {eid} {214509} (\bibinfo
  {year} {2010})},\ \Eprint {http://arxiv.org/abs/1006.2829} {arXiv:1006.2829
  [cond-mat.supr-con]} \BibitemShut {NoStop}%
\bibitem [{\citenamefont {Kwon}\ \emph {et~al.}(2004)\citenamefont {Kwon},
  \citenamefont {Yakovenko},\ and\ \citenamefont
  {Sengupta}}]{kwon_fractional_2004}%
  \BibitemOpen
  \bibfield  {author} {\bibinfo {author} {\bibfnamefont {H.-J.}\ \bibnamefont
  {Kwon}}, \bibinfo {author} {\bibfnamefont {V.~M.}\ \bibnamefont {Yakovenko}},
  \ and\ \bibinfo {author} {\bibfnamefont {K.}~\bibnamefont {Sengupta}},\
  }\href {\doibase http://dx.doi.org/10.1063/1.1789931} {\bibfield  {journal}
  {\bibinfo  {journal} {Low Temperature Physics}\ }\textbf {\bibinfo {volume}
  {30}},\ \bibinfo {pages} {613} (\bibinfo {year} {2004})}\BibitemShut
  {NoStop}%
\bibitem [{\citenamefont {Fu}\ and\ \citenamefont
  {Kane}(2009)}]{fu_josephson_2009}%
  \BibitemOpen
  \bibfield  {author} {\bibinfo {author} {\bibfnamefont {L.}~\bibnamefont
  {Fu}}\ and\ \bibinfo {author} {\bibfnamefont {C.~L.}\ \bibnamefont {Kane}},\
  }\href {\doibase 10.1103/PhysRevB.79.161408} {\bibfield  {journal} {\bibinfo
  {journal} {Phys. Rev. B}\ }\textbf {\bibinfo {volume} {79}},\ \bibinfo
  {pages} {161408} (\bibinfo {year} {2009})}\BibitemShut {NoStop}%
\bibitem [{\citenamefont {Churchill}\ \emph {et~al.}(2013)\citenamefont
  {Churchill}, \citenamefont {Fatemi}, \citenamefont {Grove-Rasmussen},
  \citenamefont {Deng}, \citenamefont {Caroff}, \citenamefont {Xu},\ and\
  \citenamefont {Marcus}}]{Churchill13}%
  \BibitemOpen
  \bibfield  {author} {\bibinfo {author} {\bibfnamefont {H.~O.~H.}\
  \bibnamefont {Churchill}}, \bibinfo {author} {\bibfnamefont {V.}~\bibnamefont
  {Fatemi}}, \bibinfo {author} {\bibfnamefont {K.}~\bibnamefont
  {Grove-Rasmussen}}, \bibinfo {author} {\bibfnamefont {M.~T.}\ \bibnamefont
  {Deng}}, \bibinfo {author} {\bibfnamefont {P.}~\bibnamefont {Caroff}},
  \bibinfo {author} {\bibfnamefont {H.~Q.}\ \bibnamefont {Xu}}, \ and\ \bibinfo
  {author} {\bibfnamefont {C.~M.}\ \bibnamefont {Marcus}},\ }\href {\doibase
  10.1103/PhysRevB.87.241401} {\bibfield  {journal} {\bibinfo  {journal} {Phys.
  Rev. B}\ }\textbf {\bibinfo {volume} {87}},\ \bibinfo {pages} {241401}
  (\bibinfo {year} {2013})}\BibitemShut {NoStop}%
\bibitem [{\citenamefont {{Finck}}\ \emph {et~al.}(2013)\citenamefont
  {{Finck}}, \citenamefont {{Van Harlingen}}, \citenamefont {{Mohseni}},
  \citenamefont {{Jung}},\ and\ \citenamefont {{Li}}}]{Van_Harlingen13}%
  \BibitemOpen
  \bibfield  {author} {\bibinfo {author} {\bibfnamefont {A.~D.~K.}\
  \bibnamefont {{Finck}}}, \bibinfo {author} {\bibfnamefont {D.~J.}\
  \bibnamefont {{Van Harlingen}}}, \bibinfo {author} {\bibfnamefont {P.~K.}\
  \bibnamefont {{Mohseni}}}, \bibinfo {author} {\bibfnamefont {K.}~\bibnamefont
  {{Jung}}}, \ and\ \bibinfo {author} {\bibfnamefont {X.}~\bibnamefont
  {{Li}}},\ }\href {\doibase 10.1103/PhysRevLett.110.126406} {\bibfield
  {journal} {\bibinfo  {journal} {Physical Review Letters}\ }\textbf {\bibinfo
  {volume} {110}},\ \bibinfo {eid} {126406} (\bibinfo {year} {2013})},\ \Eprint
  {http://arxiv.org/abs/1212.1101} {arXiv:1212.1101 [cond-mat.mes-hall]}
  \BibitemShut {NoStop}%
\bibitem [{\citenamefont {Badiane}\ \emph {et~al.}(2013)\citenamefont
  {Badiane}, \citenamefont {Glazman}, \citenamefont {Houzet},\ and\
  \citenamefont {Meyer}}]{badiane_ac_2013}%
  \BibitemOpen
  \bibfield  {author} {\bibinfo {author} {\bibfnamefont {D.~M.}\ \bibnamefont
  {Badiane}}, \bibinfo {author} {\bibfnamefont {L.~I.}\ \bibnamefont
  {Glazman}}, \bibinfo {author} {\bibfnamefont {M.}~\bibnamefont {Houzet}}, \
  and\ \bibinfo {author} {\bibfnamefont {J.~S.}\ \bibnamefont {Meyer}},\ }\href
  {\doibase 10.1016/j.crhy.2013.10.008} {\bibfield  {journal} {\bibinfo
  {journal} {Comptes Rendus Physique}\ }\textbf {\bibinfo {volume} {14}},\
  \bibinfo {pages} {840} (\bibinfo {year} {2013})}\BibitemShut {NoStop}%
\bibitem [{\citenamefont {Bretheau}\ \emph
  {et~al.}(2013{\natexlab{a}})\citenamefont {Bretheau}, \citenamefont {Girit},
  \citenamefont {Pothier}, \citenamefont {Esteve},\ and\ \citenamefont
  {Urbina}}]{bretheau2013a}%
  \BibitemOpen
  \bibfield  {author} {\bibinfo {author} {\bibfnamefont {L.}~\bibnamefont
  {Bretheau}}, \bibinfo {author} {\bibfnamefont {{\c{C}}.}~\bibnamefont
  {Girit}}, \bibinfo {author} {\bibfnamefont {H.}~\bibnamefont {Pothier}},
  \bibinfo {author} {\bibfnamefont {D.}~\bibnamefont {Esteve}}, \ and\ \bibinfo
  {author} {\bibfnamefont {C.}~\bibnamefont {Urbina}},\ }\href {\doibase
  10.1038/nature12315} {\bibfield  {journal} {\bibinfo  {journal} {Nature}\
  }\textbf {\bibinfo {volume} {499}},\ \bibinfo {pages} {312} (\bibinfo {year}
  {2013}{\natexlab{a}})}\BibitemShut {NoStop}%
\bibitem [{\citenamefont {Bretheau}\ \emph
  {et~al.}(2013{\natexlab{b}})\citenamefont {Bretheau}, \citenamefont {Girit},
  \citenamefont {Urbina}, \citenamefont {Esteve},\ and\ \citenamefont
  {Pothier}}]{Bretheau2013b}%
  \BibitemOpen
  \bibfield  {author} {\bibinfo {author} {\bibfnamefont {L.}~\bibnamefont
  {Bretheau}}, \bibinfo {author} {\bibfnamefont {{\c{C}}.~O.}\ \bibnamefont
  {Girit}}, \bibinfo {author} {\bibfnamefont {C.}~\bibnamefont {Urbina}},
  \bibinfo {author} {\bibfnamefont {D.}~\bibnamefont {Esteve}}, \ and\ \bibinfo
  {author} {\bibfnamefont {H.}~\bibnamefont {Pothier}},\ }\href {\doibase
  10.1103/PhysRevX.3.041034} {\bibfield  {journal} {\bibinfo  {journal} {Phys.
  Rev. X}\ }\textbf {\bibinfo {volume} {3}},\ \bibinfo {pages} {041034}
  (\bibinfo {year} {2013}{\natexlab{b}})}\BibitemShut {NoStop}%
\bibitem [{Note1()}]{Note1}%
  \BibitemOpen
  \bibinfo {note} {We consider a system where the phase $\varphi _0$ is
  classical. Our method does not require its quantum fluctuations, unlike, for
  example, the proposal in Ref.~\cite {Ginossar14}.}\BibitemShut {Stop}%
\bibitem [{\citenamefont {Lutchyn}\ \emph {et~al.}(2010)\citenamefont
  {Lutchyn}, \citenamefont {Sau},\ and\ \citenamefont
  {Das~Sarma}}]{lutchyn_majorana_2010}%
  \BibitemOpen
  \bibfield  {author} {\bibinfo {author} {\bibfnamefont {R.~M.}\ \bibnamefont
  {Lutchyn}}, \bibinfo {author} {\bibfnamefont {J.~D.}\ \bibnamefont {Sau}}, \
  and\ \bibinfo {author} {\bibfnamefont {S.}~\bibnamefont {Das~Sarma}},\ }\href
  {\doibase 10.1103/PhysRevLett.105.077001} {\bibfield  {journal} {\bibinfo
  {journal} {Phys. Rev. Lett.}\ }\textbf {\bibinfo {volume} {105}},\ \bibinfo
  {pages} {077001} (\bibinfo {year} {2010})}\BibitemShut {NoStop}%
\bibitem [{\citenamefont {Oreg}\ \emph {et~al.}(2010)\citenamefont {Oreg},
  \citenamefont {Refael},\ and\ \citenamefont {von
  Oppen}}]{oreg_helical_2010-1}%
  \BibitemOpen
  \bibfield  {author} {\bibinfo {author} {\bibfnamefont {Y.}~\bibnamefont
  {Oreg}}, \bibinfo {author} {\bibfnamefont {G.}~\bibnamefont {Refael}}, \ and\
  \bibinfo {author} {\bibfnamefont {F.}~\bibnamefont {von Oppen}},\ }\href
  {\doibase 10.1103/PhysRevLett.105.177002} {\bibfield  {journal} {\bibinfo
  {journal} {Phys. Rev. Lett.}\ }\textbf {\bibinfo {volume} {105}},\ \bibinfo
  {pages} {177002} (\bibinfo {year} {2010})}\BibitemShut {NoStop}%
\bibitem [{Note2()}]{Note2}%
  \BibitemOpen
  \bibinfo {note} {See Supplemental Material for details. The supplement
  includes references to~\protect \rev@citealp
  {Wang2014,potter_engineering_2011,beenakker_fermion-parity_2013,zhang_josephson_2013}.}\BibitemShut
  {Stop}%
\bibitem [{Note3()}]{Note3}%
  \BibitemOpen
  \bibinfo {note} {In the case of a NW junction, we assume that the ends of the
  wire are much farther than $\xi $ away from the junction. In the TI setup
  there are two junctions, see Fig.~\ref {fig:setup}. The outer one can be
  ignored if it is much longer than the inner one, see remark below Eq.~(\ref
  {eq:h1})}\BibitemShut {NoStop}%
\bibitem [{\citenamefont {Landau}\ and\ \citenamefont
  {Lifshitz}(1980)}]{LLVol5}%
  \BibitemOpen
  \bibfield  {author} {\bibinfo {author} {\bibfnamefont {L.}~\bibnamefont
  {Landau}}\ and\ \bibinfo {author} {\bibfnamefont {E.}~\bibnamefont
  {Lifshitz}},\ }\href@noop {} {\emph {\bibinfo {title} {Statistical
  Physics}}},\ \bibinfo {edition} {3rd}\ ed.,\ Vol.~\bibinfo {volume} {5}\
  (\bibinfo  {publisher} {Butterworth-Heinemann},\ \bibinfo {year}
  {1980})\BibitemShut {NoStop}%
\bibitem [{\citenamefont {Beenakker}(1991)}]{beenakker_universal_1991}%
  \BibitemOpen
  \bibfield  {author} {\bibinfo {author} {\bibfnamefont {C.~W.~J.}\
  \bibnamefont {Beenakker}},\ }\href {\doibase 10.1103/PhysRevLett.67.3836}
  {\bibfield  {journal} {\bibinfo  {journal} {Phys. Rev. Lett.}\ }\textbf
  {\bibinfo {volume} {67}},\ \bibinfo {pages} {3836} (\bibinfo {year}
  {1991})}\BibitemShut {NoStop}%
\bibitem [{Note4()}]{Note4}%
  \BibitemOpen
  \bibinfo {note} {The slope is $dE_\protect \mathcal {M}/d\varphi _0 = \pm
  \protect \frac {1}{2}\protect \frac {v}{L+\xi }(1+ {\protect \cal
  O}(|r_{+}|^2)) $, see Ref.~\protect \rev@citealp {Note2}.}\BibitemShut
  {Stop}%
\bibitem [{Note5()}]{Note5}%
  \BibitemOpen
  \bibinfo {note} {Equations~(\ref {eq:dEoverdphiBA}) and (\ref {eq:En}) are
  valid in the full range of $\varphi _0$ except for a narrow interval around
  $\varphi _0 =0$ where a similar avoided crossing happens.}\BibitemShut
  {Stop}%
\bibitem [{Note6()}]{Note6}%
  \BibitemOpen
  \bibinfo {note} {Lifting of degeneracies in the model for which Eq.~(\ref
  {eq:deltaphiSimple}) was derived requires, in addition to $M\not =0$, a
  non-zero chemical potential $\mu $, due to special symmetries~\cite {Note2}
  of~(\ref {eq:effH}) at the Dirac point.}\BibitemShut {Stop}%
\bibitem [{Note7()}]{Note7}%
  \BibitemOpen
  \bibinfo {note} {This selection rule is exact in a reflectionless junction,
  $\delta \varphi _n \protect \tmspace -\thinmuskip {.1667em}=\protect \tmspace
  -\thinmuskip {.1667em} 0$, where the operator $S_z$ commutes with ${H^{(0)}}$
  and $H^{(1)}$.}\BibitemShut {Stop}%
\bibitem [{\citenamefont {Klinovaja}\ and\ \citenamefont
  {Loss}(2012)}]{klinovaja_composite_2012}%
  \BibitemOpen
  \bibfield  {author} {\bibinfo {author} {\bibfnamefont {J.}~\bibnamefont
  {Klinovaja}}\ and\ \bibinfo {author} {\bibfnamefont {D.}~\bibnamefont
  {Loss}},\ }\href {\doibase 10.1103/PhysRevB.86.085408} {\bibfield  {journal}
  {\bibinfo  {journal} {Phys. Rev. B}\ }\textbf {\bibinfo {volume} {86}},\
  \bibinfo {pages} {085408} (\bibinfo {year} {2012})}\BibitemShut {NoStop}%
\bibitem [{\citenamefont {Kos}\ \emph {et~al.}(2013)\citenamefont {Kos},
  \citenamefont {Nigg},\ and\ \citenamefont {Glazman}}]{Kos13}%
  \BibitemOpen
  \bibfield  {author} {\bibinfo {author} {\bibfnamefont {F.}~\bibnamefont
  {Kos}}, \bibinfo {author} {\bibfnamefont {S.~E.}\ \bibnamefont {Nigg}}, \
  and\ \bibinfo {author} {\bibfnamefont {L.~I.}\ \bibnamefont {Glazman}},\
  }\href {\doibase 10.1103/PhysRevB.87.174521} {\bibfield  {journal} {\bibinfo
  {journal} {Phys. Rev. B}\ }\textbf {\bibinfo {volume} {87}},\ \bibinfo
  {pages} {174521} (\bibinfo {year} {2013})}\BibitemShut {NoStop}%
\bibitem [{\citenamefont {{Ginossar}}\ and\ \citenamefont
  {{Grosfeld}}(2014)}]{Ginossar14}%
  \BibitemOpen
  \bibfield  {author} {\bibinfo {author} {\bibfnamefont {E.}~\bibnamefont
  {{Ginossar}}}\ and\ \bibinfo {author} {\bibfnamefont {E.}~\bibnamefont
  {{Grosfeld}}},\ }\href {\doibase 10.1038/ncomms5772} {\bibfield  {journal}
  {\bibinfo  {journal} {Nature Communications}\ }\textbf {\bibinfo {volume}
  {5}},\ \bibinfo {eid} {4772} (\bibinfo {year} {2014})},\ \Eprint
  {http://arxiv.org/abs/1307.1159} {arXiv:1307.1159 [cond-mat.mes-hall]}
  \BibitemShut {NoStop}%
\bibitem [{\citenamefont {{Wang}}\ \emph {et~al.}(2014)\citenamefont {{Wang}},
  \citenamefont {{Gao}}, \citenamefont {{Pop}}, \citenamefont {{Vool}},
  \citenamefont {{Axline}}, \citenamefont {{Brecht}}, \citenamefont {{Heeres}},
  \citenamefont {{Frunzio}}, \citenamefont {{Devoret}}, \citenamefont
  {{Catelani}}, \citenamefont {{Glazman}},\ and\ \citenamefont
  {{Schoelkopf}}}]{Wang2014}%
  \BibitemOpen
  \bibfield  {author} {\bibinfo {author} {\bibfnamefont {C.}~\bibnamefont
  {{Wang}}}, \bibinfo {author} {\bibfnamefont {Y.~Y.}\ \bibnamefont {{Gao}}},
  \bibinfo {author} {\bibfnamefont {I.~M.}\ \bibnamefont {{Pop}}}, \bibinfo
  {author} {\bibfnamefont {U.}~\bibnamefont {{Vool}}}, \bibinfo {author}
  {\bibfnamefont {C.}~\bibnamefont {{Axline}}}, \bibinfo {author}
  {\bibfnamefont {T.}~\bibnamefont {{Brecht}}}, \bibinfo {author}
  {\bibfnamefont {R.~W.}\ \bibnamefont {{Heeres}}}, \bibinfo {author}
  {\bibfnamefont {L.}~\bibnamefont {{Frunzio}}}, \bibinfo {author}
  {\bibfnamefont {M.~H.}\ \bibnamefont {{Devoret}}}, \bibinfo {author}
  {\bibfnamefont {G.}~\bibnamefont {{Catelani}}}, \bibinfo {author}
  {\bibfnamefont {L.~I.}\ \bibnamefont {{Glazman}}}, \ and\ \bibinfo {author}
  {\bibfnamefont {R.~J.}\ \bibnamefont {{Schoelkopf}}},\ }\href {\doibase
  10.1038/ncomms6836} {\bibfield  {journal} {\bibinfo  {journal} {Nature
  Communications}\ }\textbf {\bibinfo {volume} {5}},\ \bibinfo {pages} {5836}
  (\bibinfo {year} {2014})}\BibitemShut {NoStop}%
\bibitem [{\citenamefont {Potter}\ and\ \citenamefont
  {Lee}(2011)}]{potter_engineering_2011}%
  \BibitemOpen
  \bibfield  {author} {\bibinfo {author} {\bibfnamefont {A.~C.}\ \bibnamefont
  {Potter}}\ and\ \bibinfo {author} {\bibfnamefont {P.~A.}\ \bibnamefont
  {Lee}},\ }\href {\doibase 10.1103/PhysRevB.83.184520} {\bibfield  {journal}
  {\bibinfo  {journal} {Phys. Rev. B}\ }\textbf {\bibinfo {volume} {83}},\
  \bibinfo {pages} {184520} (\bibinfo {year} {2011})}\BibitemShut {NoStop}%
\bibitem [{\citenamefont {Beenakker}\ \emph {et~al.}(2013)\citenamefont
  {Beenakker}, \citenamefont {Pikulin}, \citenamefont {Hyart}, \citenamefont
  {Schomerus},\ and\ \citenamefont {Dahlhaus}}]{beenakker_fermion-parity_2013}%
  \BibitemOpen
  \bibfield  {author} {\bibinfo {author} {\bibfnamefont {C.~W.~J.}\
  \bibnamefont {Beenakker}}, \bibinfo {author} {\bibfnamefont {D.~I.}\
  \bibnamefont {Pikulin}}, \bibinfo {author} {\bibfnamefont {T.}~\bibnamefont
  {Hyart}}, \bibinfo {author} {\bibfnamefont {H.}~\bibnamefont {Schomerus}}, \
  and\ \bibinfo {author} {\bibfnamefont {J.~P.}\ \bibnamefont {Dahlhaus}},\
  }\href {\doibase 10.1103/PhysRevLett.110.017003} {\bibfield  {journal}
  {\bibinfo  {journal} {Phys. Rev. Lett.}\ }\textbf {\bibinfo {volume} {110}},\
  \bibinfo {pages} {017003} (\bibinfo {year} {2013})}\BibitemShut {NoStop}%
\bibitem [{\citenamefont {Zhang}\ \emph {et~al.}(2013)\citenamefont {Zhang},
  \citenamefont {Zhu},\ and\ \citenamefont {Sun}}]{zhang_josephson_2013}%
  \BibitemOpen
  \bibfield  {author} {\bibinfo {author} {\bibfnamefont {S.-f.}\ \bibnamefont
  {Zhang}}, \bibinfo {author} {\bibfnamefont {W.}~\bibnamefont {Zhu}}, \ and\
  \bibinfo {author} {\bibfnamefont {Q.-f.}\ \bibnamefont {Sun}},\ }\href
  {http://stacks.iop.org/0953-8984/25/i=29/a=295301} {\bibfield  {journal}
  {\bibinfo  {journal} {Journal of Physics: Condensed Matter}\ }\textbf
  {\bibinfo {volume} {25}},\ \bibinfo {pages} {295301} (\bibinfo {year}
  {2013})}\BibitemShut {NoStop}%
\end{thebibliography}%

% The following merges the supplement into the main text


\section*{Supplementary material}

# Supplemental Material

Jukka I. Väyrynen,[1] Gianluca Rastelli,[2,3] Wolfgang Belzig,[3] and Leonid I. Glazman[1]

[1]*Department of Physics, Yale University, New Haven, CT 06520, USA*
[2]*Zukunftskolleg, Universität Konstanz, D-78457 Konstanz, Germany*
[3]*Fachbereich Physik, Universität Konstanz, D-78457 Konstanz, Germany*

In this supplemental material we show in detail the derivations that were omitted in the main text.

## SM1. DERIVATION OF THE CURRENT OPERATOR

To model the S-TI-S (or S-NW-S) junction, we assume that the left ($x < 0$) and right ($x > L$) halves of the one-dimensional edge are tunnel-coupled (with tunneling Hamiltonians $H_{LT}$ and $H_{RT}$) to the respective leads. The junction comprises of the segment $0 < x < L$, where $L$ is the distance between the leads, see Fig. 1 of the main text. The operator of tunneling current into the right lead is

$$I = e\frac{i}{\hbar}[H_{RT}, N_{RL}].$$ (S1)

Since $H_{RT}$ conserves the total number of electrons, it commutes with $N_{RL} + N_R$, where $N_{RL}$ and $N_R$ are the respective electron number operators for the right lead and the part of the edge in tunnel-contact with the lead [S1]. The conservation law allows us to write Eq. (S1) as $I = -e(i/\hbar)[H_{RT}, N_R]$. Since we are interested in the physics of the junction bound states, it is beneficial to move to a low-energy description that includes only the edge degrees of freedom. This is possible because the superconducting leads are gapped; the description is valid at energies less than the pairing gaps of the leads, $E \ll \Delta$. The current operator $I$ projected to the low-energy degrees of freedom is easiest found by writing it first in a form that does not explicitly contain the lead operators. Denoting $H'$ the full (bare) Hamiltonian of the edge, and $H'_K$ its kinetic energy term, we can write $[H_{RT}, N_R] = [(H' - H'_K), N_R]$ (since the only terms in $H'$ that do not commute with $N_R$ are $H'_K$ and $H_{RT}$). This yields

$$I = -e(i/\hbar)[(H' - H'_K), N_R].$$ (S2)

Now we are ready to project $I$ to the low-energy subspace. At low energies $E \ll \Delta$, *single* electrons from the edge cannot tunnel into the leads. Pairs of electrons, however, can tunnel, and this gives rise to an effective pairing between them. When moving to the low-energy description, the aforementioned tunneling term $H_{RT}$ is replaced by a particle non-conserving pairing Hamiltonian $H_{\Delta,R}$, given by

$$H_{\Delta,R} = \frac{1}{2}\Delta_0 \int_L^{L+a} dx\, \Psi^\dagger(\tau_x \cos\varphi - \tau_y \sin\varphi)\Psi,$$ (S3)

where $a$ is the length of the right proximitized region (see Fig. 1 of the main text), and $\varphi$ is the relative phase between the leads (we set the phase of the left superconductor to zero). The magnitude of the proximity induced gap at energies $E \ll \Delta$ is $\Delta_0 = \frac{1}{2}\gamma\Delta/(\Delta + \frac{1}{2}\gamma)$ [S2], where $\gamma$ is the tunneling rate between the edge and the lead in absence of $\Delta$. In Eq. (S3) we introduced the spinor notation $\Psi^\dagger = (\psi_\uparrow^\dagger, \psi_\downarrow^\dagger, \psi_\downarrow, -\psi_\uparrow)$, where $\psi_{\uparrow(\downarrow)}^\dagger$ creates a right (left) moving electron on the edge. The Pauli matrices $\tau_{x,y,z}$ act in the particle-hole space. In the spinor notation $N_R = \int_L^{L+a} dx\, \Psi^\dagger \tau_z \Psi/2$. Projecting the current operator $I$ to low energies, we can replace $(H' - H'_K) \to H_{\Delta,R}$ in Eq. (S2). Commuting $N_R$ with $H_{\Delta,R}$, we find that the projected current operator can be written as

$$I = 2e\frac{1}{\hbar}\frac{\partial H}{\partial \varphi},$$ (S4)

since $H_{\Delta,R}$ is the only $\varphi$-dependent term in $H$.

To irradiate the junction with microwaves, one applies a weak time-dependent voltage $V(t)$ to, say, the right lead, where we measure the current. (The specific relation between $V(t)$ and the electric field amplitude of the microwaves depends on a concrete device geometry [S3].) The effect of the resulting bias voltage is to make the low energy Hamiltonian $H$ time-dependent: the phase difference $\varphi$ in Eq. (S3) has to be replaced by $\varphi(t) = \varphi_0 + \delta\varphi(t)$, where $\varphi_0$ is the phase difference in absence of bias and $\delta\varphi(t)$ satisfies the Josephson relation $\delta\dot\varphi(t) = 2eV(t)/\hbar$. The replacement can be derived by doing a time-dependent gauge transformation that removes the bias from the leads. The transformation gives rise to a time-dependent phase in the, say, right tunneling amplitudes and, at low energies, in the order parameter in Eq. (S3). Developing (S3) to first order in $\delta\dot\varphi(t)$ leads to Eq. (6) of the main text.

## SM2. THE FULL EXPRESSION FOR THE ADMITTANCE

The admittance $Y(\omega)$ is found by using the standard Kubo formula with $H^{(1)}$ [Eq. (6) of the main text] as the perturbation. We find ($\omega^+ = \omega + 0i$),

$$
\begin{aligned}
Y(\omega) = & \frac{i}{\omega^+ L_J} \\
& + \frac{8ie^2}{\omega^+} \sum_{E_n > E_m \geq 0} \left\{ \frac{|\mathcal{H}^{(1)}_{n;m}|^2 (E_n - E_m)}{(\omega^+)^2 - (E_n - E_m)^2}[f(E_m) - f(E_n)] + \frac{|\mathcal{H}^{(1)}_{n;-m}|^2 (E_n + E_m)}{(\omega^+)^2 - (E_n + E_m)^2}[1 - f(E_m) - f(E_n)] \right\}.
\end{aligned}
\tag{S5}
$$

Here $f(E) = 1/(e^{E/T} + 1)$ is the Fermi function. The transitions in (S5) conserve the fermion (and quasiparticle) number parity: the first term in the brackets arises from transitions between two different quasiparticle states, while the second term accounts for creation/annihilation of a pair of quasiparticles. There are no terms that create an odd number of quasiparticles. Also, there are no terms with $E_m = E_n$ since none of the levels are degenerate in a generic disordered junction and $\mathcal{H}^{(1)}_{n;-n} = 0$ by Pauli exclusion principle. The first term in Eq. (S5) represents the conventional inductive response of a Josephson junction. The inverse inductance $1/L_J = 2e\partial_{\varphi_0} I_J$ is a derivative of the Josephson current, $I_J(\varphi_0) = \langle I \rangle_0$, with the current operator defined in Eq. (5) of the main text, and $\langle \ldots \rangle_0$ denoting the averaging over an equilibrium density matrix with the Hamiltonian (3) of the main text.

Equation (8) of the main text is obtained from Eq. (S5) by taking the real part and accounting only for the positive frequency transitions involving the lowest level, $E_m = E_{\mathcal{M}}$.

## SM3. THE EQUATION FOR THE BOUND STATE ENERGY SPECTRUM

The bound state energy spectrum is calculated by using continuity of the eigenstate wave function $\Phi_E(x)$ at the two interfaces, $x = 0$, $L$. This leads to the condition (see Sec. SM3 A and Ref. S4)

$$
\det[1 - S_N(E)S_A(E)] = 0,
\tag{S6}
$$

which determines the energy spectrum. Here $S_N$ and $S_A$ are respectively the normal and Andreev scattering matrices (defined in Sec. SM3 A). The former describes the reflection in the junction ($0 < x < L$), while the latter accounts for the Andreev reflection at the two interfaces at $x = 0$, $L$. We assume hereafter that there is backscattering only in the junction, $i.e.$, the Zeeman field is vanishing outside it, $M(x) = 0$ for $x < 0$ and $x > L$ [S5]. In particle-hole space the two matrices are

$$
S_N(E) = \begin{pmatrix} S_+(E) & 0 \\ 0 & S_-(E) \end{pmatrix}, \quad S_A(E) = e^{-i\alpha(E)} \begin{pmatrix} 0 & r_A \\ r_A^* & 0 \end{pmatrix},
\tag{S7}
$$

where $e^{-i\alpha(E)} = \Delta_0/(E + i\kappa v)$, and $\kappa v = \sqrt{\Delta_0^2 - E^2}$ is real for bound states. The components in spin-space are

$$
S_\pm(E) = \begin{pmatrix} r_\pm(E) & d'_\pm(E) \\ d_\pm(E) & r'_\pm(E) \end{pmatrix}, \quad r_A = \begin{pmatrix} 1 & 0 \\ 0 & e^{i\varphi_0} \end{pmatrix}.
\tag{S8}
$$

The scattering amplitudes $r_\pm^{(\prime)}(E)$, $d_\pm^{(\prime)}(E)$ in terms of the junction parameters are given in Sec. SM4. Hereinafter, we omit their energy argument in the notation, unless there is a risk of confusion. Inserting Eqs. (S7) and (S8) into Eq. (S6), and using the unitarity of $S_\pm$ yields the equation for the spectrum,

$$
\begin{aligned}
& e^{2i\alpha(E)} + (d_+ d'_- - r_+ r'_+)(d_- d'_- - r_- r'_-)e^{-2i\alpha(E)} \\
& - d_+ d'_- e^{-i\varphi_0} - d'_+ d_- e^{i\varphi_0} - r_+ r_- - r'_+ r'_- = 0.
\end{aligned}
\tag{S9}
$$

The equation (S9) is difficult to solve in full generality, and its tractable limit will be considered in Sec. SM6. However, already at this stage we note that this equation has a solution $E_{\mathcal{M}}(\varphi_0)$ that vanishes at $\varphi_0 = \pi$. This can be seen by using in Eq. (S9) the following two properties of the scattering matrices $S_\pm$ [which follow from two symmetries of the Hamiltonian (3) of the main text]. First, from particle-hole symmetry, one has $S_-(E) = -\sigma_z S_+^*(-E)\sigma_z$ [see Eq. (S42); $*$ denotes complex conjugation]. Second, from the invariance of junction Hamiltonian under the combined reversal of time and magnetic field [$M(x) \to -M(x)$], one finds $d_\pm = d'_\pm$ [see Eq. (S44)]. These two properties of the scattering matrix are derived in Sec. SM4 B. Using the two properties as well as the unitarity of the scattering matrix, $S_\pm^\dagger S_\pm = 1$, it is straightforward to check that Eq. (S9) has a solution $E_{\mathcal{M}} = 0$ at $\varphi_0 = \pi$.

By expanding Eq. (S9) in $\varphi_0 - \pi \ll 1$, we can evaluate the slope $\partial_{\varphi_0} E_{\mathcal{M}}(\pi)$ of the lowest Andreev level. We find (see Sec. SM3 B),

$$\partial_{\varphi_0} E_{\mathcal{M}}(\pi) = \pm \frac{1}{2} \frac{|d_+|^2}{\sqrt{\left|\partial_E d_+ + i d_+ \Delta_0^{-1}\right|^2 - \left|\partial_E (d_+ r_+^*)\right|^2}}\,, \tag{S10}$$

where the scattering amplitudes on the right-hand-side are evaluated at $E = 0$ (in which case $|d_+| = |d_-|$). Let us consider two simple limits of Eq. (S10). First, when $\Delta_0$ is much smaller than the energy scale at which the scattering amplitudes vary, we recover $\partial_{\varphi_0} E_{\mathcal{M}}(\pi) = \pm |d_+| \Delta_0/2$ which was already found in Ref. [S5] for a junction of length $L \ll \xi$ with $M(x) = M\Theta(x)\Theta(L - x)$. The second example is that of a weak magnetic field $M(x)$. We can then use the Born approximation (BA) for the scattering matrix. The reflection amplitudes are small, $|r_\pm^{(\prime)}| \ll 1$, and the second term in the denominator of Eq. (S10) can be neglected. In the model of Eq. (3) in the main text, the BA is valid when $\int_0^L dx M(x)/v \ll 1$, and amounts to setting $d_+ = e^{i\delta_+}$ with $\partial_E \delta_+ = L/v$ (see Sec. SM4). Equation (S10) then becomes

$$\partial_{\varphi_0} E_{\mathcal{M}}(\pi) = \pm \frac{1}{2} \frac{v}{L + \xi} + \mathcal{O}(|r_+|^2)\,, \tag{S11}$$

leading to Eq. (9) of the main text (see also the footnote 30 there). The terms $\mathcal{O}(|r_+|^2)$ are given in Eq. (S88) in Sec. SM6 E. Finally, we wish to emphasize that Eqs. (S10) and (S11) are true for a generic disordered junction with any scalar disorder potential $V(x)$ that does not exceed the TI bulk band gap. The problem is tractable, because of the topological protection of the helical states against scalar disorder – the potential $V(x)$ does not backscatter electrons and the junction is reflectionless in absence of $M(x)$.

### A. Derivation of Eq. (S6) and the wave function boundary values

We write the wave function as $\Phi = (u_\uparrow, u_\downarrow, v_\downarrow, v_\uparrow)^T$, suppressing the energy labels in this section. The scattering matrices for the two superconducting regions ($x < 0$ and $x > L$) are defined as

$$\begin{pmatrix} u_\downarrow(-\infty) \\ u_\uparrow(0) \\ v_\downarrow(-\infty) \\ v_\uparrow(0) \end{pmatrix} = S_{A,L} \begin{pmatrix} u_\uparrow(-\infty) \\ u_\downarrow(0) \\ v_\uparrow(-\infty) \\ v_\downarrow(0) \end{pmatrix}, \quad \begin{pmatrix} u_\downarrow(L) \\ u_\uparrow(\infty) \\ v_\downarrow(L) \\ v_\uparrow(\infty) \end{pmatrix} = S_{A,R} \begin{pmatrix} u_\uparrow(L) \\ u_\downarrow(\infty) \\ v_\uparrow(L) \\ v_\downarrow(\infty) \end{pmatrix}. \tag{S12}$$

where the incoming states are on the right. (We shall shortly require $\Phi(\pm\infty) = 0$ since we are considering bound states.) The Andreev scattering matrix for the left/right "semi-infinite" superconducting region is

$$S_{A,L/R} = e^{-i\alpha(E)} \begin{pmatrix} 0 & e^{i\varphi_{L/R}} \begin{pmatrix} 1 & 0 \\ 0 & 1 \end{pmatrix} \\ e^{-i\varphi_{L/R}} \begin{pmatrix} 1 & 0 \\ 0 & 1 \end{pmatrix} & 0 \end{pmatrix}, \quad e^{-i\alpha(E)} = \frac{\Delta_0}{E + iv\kappa}, \quad \kappa = \frac{1}{v}\sqrt{\Delta_0^2 - E^2}, \tag{S13}$$

and is derived in Sec. SM4 A. For bound states $\Phi(\pm\infty) = 0$, and Eq. (S12) can be replaced by the single equation

$$\begin{pmatrix} u_\uparrow(0) \\ u_\downarrow(L) \\ v_\uparrow(0) \\ v_\downarrow(L) \end{pmatrix} = S_A \begin{pmatrix} u_\downarrow(0) \\ u_\uparrow(L) \\ v_\downarrow(0) \\ v_\uparrow(L) \end{pmatrix}, \quad S_A = e^{-i\alpha(E)} \begin{pmatrix} 0 & r_A \\ r_A^* & 0 \end{pmatrix}, \quad r_A = \begin{pmatrix} 1 & 0 \\ 0 & e^{i\varphi_0} \end{pmatrix}, \tag{S14}$$

where we set $\varphi_L = 0$, $\varphi_R = \varphi_0$.

In the junction, $0 < x < L$, we have the normal state scattering relation

$$\begin{pmatrix} u_\downarrow(0) \\ u_\uparrow(L) \\ v_\downarrow(0) \\ v_\uparrow(L) \end{pmatrix} = S_N \begin{pmatrix} u_\uparrow(0) \\ u_\downarrow(L) \\ v_\uparrow(0) \\ v_\downarrow(L) \end{pmatrix}, \quad S_N = \begin{pmatrix} S_+ & 0 \\ 0 & S_- \end{pmatrix}, \quad S_\pm = \begin{pmatrix} r_\pm & d_\pm' \\ d_\pm & r_\pm' \end{pmatrix}. \tag{S15}$$

where $r_{+/-}^{(\prime)}$ and $d_{+/-}^{(\prime)}$ are the reflection and transmission amplitudes for electrons/holes entering the junction from the left (right). The eight equations in (S14) and (S15) allow us to express all the components of $\Phi(0)$ and $\Phi(L)$ in

terms of a single component, say $v_\uparrow(L)$, which can be taken to be the wave function normalization constant. One of the eight equations remains unused – this is Eq. (S6) which determines the energy spectrum. Explicitly, it reads,

$$\left(d_- e^{i\alpha(E)} - d_+(d_- d'_- - r_- r'_-)e^{-i\alpha(E)}e^{-i\varphi_0}\right)\left(d_+ e^{i\alpha(E)} - d_-(d_+ d'_+ - r_+ r'_+)e^{-i\alpha(E)}e^{i\varphi_0}\right)$$
$$- \left(d_+ r_- + d_- r'_+ e^{i\varphi_0}\right)\left(d_- r_+ + d_+ r'_- e^{-i\varphi_0}\right) = 0 \,. \tag{S16}$$

Using unitarity of the scattering matrices, $S_\pm^\dagger S_\pm = 1$, in (S16) leads to Eq. (S9).

The wave function components are

$$u_\uparrow(0) = \frac{r'_+(d_- d'_- - r_- r'_-)e^{-i\alpha(E)} + r_- e^{i\alpha(E)}}{d_- e^{i\alpha(E)} - d_+(d_- d'_- - r_- r'_-)e^{-i\alpha(E)}e^{-i\varphi_0}}e^{-i\alpha(E)}v_\uparrow(L) \,, \tag{S17a}$$

$$v_\downarrow(0) = \frac{r'_+(d_- d'_- - r_- r'_-)e^{-i\alpha(E)} + r_- e^{i\alpha(E)}}{d_- e^{i\alpha(E)} - d_+(d_- d'_- - r_- r'_-)e^{-i\alpha(E)}e^{-i\varphi_0}}v_\uparrow(L) \,, \tag{S17b}$$

$$u_\downarrow(0) = \frac{r'_-(d_+ d'_+ - r_+ r'_+)e^{-i\alpha(E)} + r_+ e^{i\alpha(E)}}{d_- r_+ + d_+ r'_- e^{-i\varphi_0}}v_\uparrow(L) \,, \tag{S17c}$$

$$v_\uparrow(0) = \frac{r'_-(d_+ d'_+ - r_+ r'_+)e^{-i\alpha(E)} + r_+ e^{i\alpha(E)}}{d_- r_+ + d_+ r'_- e^{-i\varphi_0}}e^{-i\alpha(E)}v_\uparrow(L) \,, \tag{S17d}$$

and

$$u_\uparrow(L) = \frac{d_+ r_- + d_- r'_+ e^{i\varphi_0}}{d_- e^{i\alpha(E)} - d_+(d_- d'_- - r_- r'_-)e^{-i\alpha(E)}e^{-i\varphi_0}}v_\uparrow(L) \,, \tag{S17e}$$

$$v_\downarrow(L) = \frac{d_+ r_- + d_- r'_+ e^{i\varphi_0}}{d_- e^{i\alpha(E)} - d_+(d_- d'_- - r_- r'_-)e^{-i\alpha(E)}e^{-i\varphi_0}}e^{-i\alpha(E)}e^{-i\varphi_0}v_\uparrow(L) \,, \tag{S17f}$$

$$u_\downarrow(L) = e^{-i\alpha(E)}e^{i\varphi_0}v_\uparrow(L) \,. \tag{S17g}$$

The spatial dependence of $\Phi(x)$ can be found by acting on, say, $\Phi(L)$ with the transfer matrix, see Eq. (S31) in Sec. SM4.

## B. Derivation of Eq. (S10) from Eq. (S9)

It is useful to divide Eq. (S9) into two parts,

$$f_1(E) + f_2(E, \varphi_0) = 0 \,, \tag{S18}$$

where $f_1$ does not contain explicit $\varphi_0$-dependence,

$$f_1(E) = e^{2i\alpha(E)} + (d_+^2 - r_+ r'_+)(d_-^2 - r_- r'_-)e^{-2i\alpha(E)} - r_+ r_- - r'_+ r'_- \,, \tag{S19}$$
$$f_2(E, \varphi_0) = -2d_+ d_- \cos\varphi_0 \,.$$

We have used here the property $d = d'_\pm$, see Eq. (S44). Using $d/d\varphi_0 = \partial_{\varphi_0} E \partial_E + \partial_{\varphi_0}$, the first derivative of Eq. (S18) yields ($E' = \partial_{\varphi_0} E$)

$$E' \partial_E f_1(E) + E' \partial_E f_2(E, \varphi_0) + \partial_{\varphi_0} f_2(E, \varphi_0) = 0 \,, \tag{S20a}$$

Note that at $\varphi_0 = \pi$ we have $\partial_{\varphi_0} f_2 = 0$ and thus $\partial_E f_1 = -\partial_E f_2$ from Eq. (S20a). Taking the second derivative of Eq. (S18) yields

$$E'^2\left(\partial_E^2 f_1(E) + \partial_E^2 f_2(E, \varphi_0)\right) + E''\left(\partial_E f_1(E) + \partial_E f_2(E, \varphi_0)\right) + E'\partial_{\varphi_0}\partial_E f_1(E) + 2E'\partial_E\partial_{\varphi_0}f_2(E, \varphi_0) + \partial_{\varphi_0}^2 f_2(E, \varphi_0) = 0 \,. \tag{S20b}$$

Evaluating this at $\varphi_0 = \pi$ and using Eq. (S20a) leads to

$$E'^2\left(\partial_E^2 f_1(E) + \partial_E^2 f_2(E, \varphi_0)\right) + \partial_{\varphi_0}^2 f_2(E, \varphi_0) = 0 \,. \tag{S21}$$

From Eq. (S18) we have $f_1 = -f_2 = \partial_{\varphi_0}^2 f_2$, and thus

$$(\partial_\varphi E)^2 = \left[\frac{f_2(E, \pi)}{\partial_E^2 (f_1(E) + f_2(E, \pi))}\right]_{E=0}.$$
(S22)

Next we will evaluate the derivatives in the denominator. From particle-hole symmetry, Eq. (S42), we have the following properties valid at zero energy: $\partial_E^{(n)} d_+ = (-1)^n \partial_E^{(n)} d_-^*$ and $\partial_E^{(n)} r_+^{(\prime)} = (-1)^{n+1} \partial_E^{(n)} r_-^{(\prime)*}$ for the $n$th derivative, $n = 0$ meaning no derivative. Using Eq. (S19) and the properties $\partial_E e^{\pm 2i\alpha(E)} = \pm 2i/\Delta_0$ and $\partial_E^2 e^{\pm 2i\alpha(E)} = 4/\Delta_0^2$, we find

$$\partial_E^2 (f_1(E) + f_2(E, \varphi_0)) = \frac{8}{\Delta_0^2} + \frac{4}{|d_+|^2} \left|\frac{\partial_E(d_+ d_-)}{d_+}\right|^2 + \frac{8}{|d_+|^2} \frac{(-i)}{\Delta_0} \partial_E(d_+ d_-) - \partial_E^2[r_+ r_- + r_+' r_-' - 2d_+ d_-]$$

$$= \frac{8}{|d_+|^2} \left(\left|\partial_E d_+ + \frac{id_+}{\Delta_0}\right|^2 - |\partial_E d_+|^2 + \frac{1}{2} \left|\frac{\partial_E(d_+ d_-)}{d_+}\right|^2 - |d_+|^2 \frac{1}{8} \partial_E^2[r_+ r_- + r_+' r_-' - 2d_+ d_-]\right).$$
(S23)

We can get rid of the second order derivatives in the last term. First, we note that

$$\partial_E^2(d_- d_+) = -2|\partial_E d_+|^2 + 2\text{Re}d_+^* \partial_E^2 d_+,$$

$$\partial_E^2(r_+^{(\prime)} r_-^{(\prime)}) = 2|\partial_E r_+^{(\prime)}|^2 - 2\text{Re}r_+^{(\prime)*} \partial_E^2 r_+^{(\prime)},$$
(S24a)

and also

$$\text{Re}d_+^* \partial_E^2 d_+ = -|\partial_E d_+|^2 - |\partial_E r_+|^2 - \text{Re}r_+^* \partial_E^2 r_+$$

$$= -|\partial_E d_+|^2 - |\partial_E r_+'|^2 - \text{Re}r_+'^* \partial_E^2 r_+',$$
(S24b)

as can be found by differentiating the unitarity condition $|d_+|^2 + |r_+'|^2 = |d_+|^2 + |r_+|^2 = 1$. Combining the above two equations gives,

$$\partial_E^2(d_- d_+) = -4|\partial_E d_+|^2 - |\partial_E r_+|^2 - |\partial_E r_+'|^2 - \text{Re}r_+^* \partial_E^2 r_+ - \text{Re}r_+'^* \partial_E^2 r_+'$$
(S25a)

$$\partial_E^2(r_+^{(\prime)} r_-^{(\prime)}) = 2|\partial_E r_+^{(\prime)}|^2 - 2\text{Re}r_+^{(\prime)*} \partial_E^2 r_+^{(\prime)}$$
(S25b)

The second order derivatives cancel in Eq. (S23), and we are left with

$$\partial_E^2 (f_1(E) + f_2(E, \varphi_0)) = \frac{8}{|d_+|^2} \left(\left|\partial_E d_+ + \frac{id_+}{\Delta_0}\right|^2 - |\partial_E d_+|^2 + \frac{1}{2} \left|\frac{\partial_E(d_+ d_-)}{d_+}\right|^2 - \frac{1}{2}|d_+|^2[|\partial_E r_+|^2 + |\partial_E r_+'|^2 + 2|\partial_E d_+|^2]\right).$$
(S26)

We will next show that the r.h.s. of the above equation can be simplified,

$$-|\partial_E d_+|^2 + \frac{1}{2} \left|\frac{\partial_E(d_+ d_-)}{d_+}\right|^2 - \frac{1}{2}|d_+|^2[|\partial_E r_+|^2 + |\partial_E r_+'|^2 + 2|\partial_E d_+|^2] = -|\partial_E(d_-^* r_+)|^2.$$
(S27)

Using the unitarity of scattering matrix, $r_+' = -d_+ r_+^*/d_+^*$, we have

$$|\partial_E r_+'|^2 = |\partial_E r_+|^2 - 4\frac{1}{|d_+|^2}\text{Im}d_+ \partial_E d_+^* \text{Im}r_+ \partial_E r_+^* + 4\frac{|r_+|^2}{|d_+|^4}(\text{Im}d_+ \partial_E d_+^*)^2$$
(S28)

where we used $\frac{d_+}{d_+^*}\partial_E\left(\frac{d_+^*}{d_+}\right) = 2i\frac{1}{|d_+|^2}\text{Im}\frac{\partial_E d_+^*}{d_+^*}$ in the second and third terms. Finally inserting above equation into the left-hand-side of Eq. (S27) and using $\partial_E(d_+ d_-) = 2i\text{Im}d_+ \partial_E d_+$, we find

$$-|\partial_E d_+|^2 + \frac{1}{2} \left|\frac{\partial_E(d_+ d_-)}{d_+}\right|^2 - \frac{1}{2}|d_+|^2[|\partial_E r_+|^2 + |\partial_E r_+'|^2 + 2|\partial_E d_+|^2]$$

$$= -|\partial_E d_+|^2 + 2(\text{Im}d_+ \partial_E d_+^*)^2 - |d_+|^2|\partial_E r_+|^2 + 2\text{Im}d_+ \partial_E d_+^* \text{Im}r_+ \partial_E r_+^* - |d_+|^2|\partial_E d_+|^2$$

$$= -|r_+|^2|\partial_E d_+|^2 - |d_+|^2|\partial_E r_+|^2 + 2\left(\text{Re}(d_+ \partial_E d_+^*)\text{Re}(r_+^* \partial_E r_+) - \text{Im}d_+ \partial_E d_+^* \text{Im}r_+^* \partial_E r_+\right)$$

$$= -|r_+|^2|\partial_E d_+|^2 - |d_+|^2|\partial_E r_+|^2 + 2\text{Re}(d_+ \partial_E d_+^*)(r_+^* \partial_E r_+)$$

$$= -|\partial_E(d_-^* r_+)|^2.$$
(S29)

In the third line we used the complex number identity $(\mathrm{Im}\,w)^2 = |w|^2 - (\mathrm{Re}\,w)^2$, and also $\mathrm{Re}(d_+ \partial_E d_+^*) = -\mathrm{Re}(r_+^* \partial_E r_+)$ from unitarity.

With the above identity, we can write

$$\partial_E^2 \left(f_1(E) + f_2(E, \varphi_0)\right) = \frac{8}{|d_+|^2} \left( \left| \partial_E d_+ + \frac{i d_+}{\Delta_0} \right|^2 - |\partial_E (d_-^* r_+)|^2 \right) . \tag{S30}$$

Inserting this into Eq. (S22) leads to Eq. (S10).

## SM4.  THE SCATTERING MATRICES

The Andreev and normal scattering matrices $S_A$ and $S_N$ were already defined in Sec. SM3, in Eqs. (S12) and (S15). In this section we express them in terms of the microscopic parameters of the model Eq. (3) in the main text. This is done by using the transfer matrix. The Andreev and normal scattering are discussed in Secs. SM4 A and SM4 B, respectively. Finally, in Sec. SM4 C we derive the normal reflection amplitudes in the Born approximation for a weakly reflecting junction. This result will be put to use in Sec. SM6.

Since $\mathcal{H}^{(0)}(x)$, Eq. (3) of the main text, has only one spatial derivative, its eigenstates can be conveniently written with the help of the transfer matrix. An eigenstate $\Phi_E$ of energy $E$ can be written as

$$\Phi_E(x) = T(x, x_0; E)\Phi_E(x_0), \tag{S31}$$

where the transfer matrix is

$$T(x, x_0; E) = \mathcal{T} \exp \int_{x_0}^{x} dx' \frac{i}{v} \tau_z \sigma_z \left(E + [\mu - V(x')]\tau_z - M(x')\sigma_x - \Delta_0(x')[\tau_x \cos \varphi(x') - \tau_y \sin \varphi(x')]\right), \tag{S32}$$

and the ordered exponential is defined as,

$$\mathcal{T} \exp \int_{x_0}^{x} dx' f(x') = \lim_{N \to \infty} e^{f(x_N)\Delta x} e^{f(x_{N-1})\Delta x} \dots e^{f(x_0)\Delta x}, \quad \Delta x = \frac{x - x_0}{N}, \quad x_n = x_0 + n\Delta x, \tag{S33}$$

and satisfies $\partial_x \mathcal{T} \exp \int_{x_0}^{x} dx' f(x') = f(x)\mathcal{T} \exp \int_{x_0}^{x} dx' f(x')$.

From the transfer matrices $T(-\infty, 0)$, $T(0, L)$, and $T(L, \infty)$, we can read off the scattering matrices $S_{A,L}$, $S_N$, and $S_{A,R}$, introduced in Sec. SM3, Eqs. (S12) and (S15). (Hereafter, we shall mostly suppress the explicit $E$ in the arguments of the transfer matrix and scattering matrix.) We will start by considering the superconducting regions, $x < 0$ and $x > L$.

### A.  Transfer matrices outside the junction

We assume that $M(x') = 0$ in the superconducting regions. The transfer matrix (S32) is then diagonal in spin-indices. The term $i\sigma_z[\mu - V(x')]/v$ in the exponential in (S32) commutes with the other terms, and can be pulled out to factorize the transfer matrix. Also, in our model (3) of the main text, $\Delta_0$ and $\varphi$ are piecewise constants so that

$$T(x, x_0) = e^{\frac{i}{v}\sigma_z \int_{x_0}^{x} [\mu - V(x')] dx'} e^{\sigma_z(\hat{n} \cdot \boldsymbol{\tau})\kappa(x - x_0)}, \quad \hat{n} = \frac{\Delta_0(\mathbf{y}\cos\varphi + \mathbf{x}\sin\varphi) + iE\mathbf{z}}{\kappa v}, \quad \kappa v = \sqrt{\Delta_0^2 - E^2}, \tag{S34}$$

where either $x, x_0 < 0$ (in which case we set $\varphi = 0$ in $\hat{n}$), or $x, x_0 > L$ (in which case $\varphi = \varphi_0$). Using Eq. (S34) in Eq. (S31) with $(x, x_0) \to (0, -\infty)$, and $(x, x_0) \to (\infty, L)$, leads to Eqs. (S12), (S13).

Using the identity

$$e^{\sigma_z(\hat{n} \cdot \boldsymbol{\tau})\kappa(x - x_0)} = \frac{1}{2} \left[ (1 + \sigma_z(\hat{n} \cdot \boldsymbol{\tau})) e^{\kappa(x - x_0)} + (1 - \sigma_z(\hat{n} \cdot \boldsymbol{\tau})) e^{-\kappa(x - x_0)} \right], \tag{S35}$$

in Eq. (S34), and requiring that $\Phi_E(\pm\infty) = 0$ we find the equations

$$(1 + \sigma_z(\hat{n} \cdot \boldsymbol{\tau}))\Phi_E(L) = 0, \quad (1 - \sigma_z(\hat{n} \cdot \boldsymbol{\tau}))\Phi_E(0) = 0, \tag{S36}$$

which lead to Eq. (S14). Combining the above three equations in Eq. (S31) shows that the wave functions decay exponentially outside the junction,

$$\Phi_E(x) = \begin{cases} e^{\frac{i}{v}\sigma_z \int_L^x [\mu - V(x')] dx'} e^{-\kappa(x - L)}\Phi_E(L), & x > L, \\ e^{\frac{i}{v}\sigma_z \int_0^x [\mu - V(x')] dx'} e^{\kappa x}\Phi_E(0), & x < 0. \end{cases} \tag{S37}$$

The decay length is $\kappa^{-1}$ and approximately equal to $\xi$ at energies $E \ll \Delta_0$.

## B. Transfer matrices inside the junction

In the junction we have $\Delta_0(x') = 0$ in Eq. (S32). The transfer matrix $T(L,0)$ is therefore diagonal in the particle-hole components. The diagonal elements are explicitly

$$T_\pm = \mathcal{T} \exp \int_0^L dx' \frac{\pm i}{v} \sigma_z \left( E \pm [\mu - V(x')] - M(x')\sigma_x \right).$$ (S38)

Here $T_{+/-}$ corresponds to the electron/hole transfer matrix. For brevity, we suppress their position arguments in this section. To relate the components of the scattering matrix to those of the transfer matrix, we write Eq. (S31) across the junction:

$$\begin{pmatrix} u_\uparrow(L) \\ u_\downarrow(L) \\ v_\downarrow(L) \\ v_\uparrow(L) \end{pmatrix} = \begin{pmatrix} T_+ & 0 \\ 0 & T_- \end{pmatrix} \begin{pmatrix} u_\uparrow(0) \\ u_\downarrow(0) \\ v_\downarrow(0) \\ v_\uparrow(0) \end{pmatrix}, \quad T_\pm = \begin{pmatrix} T_{\pm\uparrow\uparrow} & T_{\pm\uparrow\downarrow} \\ T_{\pm\downarrow\uparrow} & T_{\pm\downarrow\downarrow} \end{pmatrix}.$$ (S39)

Comparing Eq. (S39) to the definition of the junction scattering matrix $S_N$, Eq. (S15), we find

$$\begin{pmatrix} r_+ & d'_+ \\ d_+ & r'_+ \end{pmatrix} = \begin{pmatrix} -\frac{1}{T_{+\downarrow\downarrow}} T_{+\downarrow\uparrow} & \frac{1}{T_{+\downarrow\downarrow}} \\ \frac{1}{T_{+\uparrow\uparrow}} & \frac{1}{T_{+\downarrow\downarrow}} T_{+\uparrow\downarrow} \end{pmatrix}, \quad \begin{pmatrix} r_- & d'_- \\ d_- & r'_- \end{pmatrix} = \begin{pmatrix} -\frac{1}{T_{-\uparrow\uparrow}} T_{-\uparrow\downarrow} & \frac{1}{T_{-\uparrow\uparrow}} \\ \frac{1}{T_{-\downarrow\downarrow}} & \frac{1}{T_{-\uparrow\uparrow}} T_{-\downarrow\uparrow} \end{pmatrix}.$$ (S40)

In deriving these expressions, we used the pseudounitarity, $T_\pm^\dagger \sigma_z T_\pm = \sigma_z$, of the electron and hole transfer matrices. One can check that the pseudounitarity implies unitarity, $S_\pm^\dagger S_\pm = 1$, of the scattering matrices.

Particle-hole transformation $P = \sigma_y \tau_y K$ ($K$ denotes complex conjugation) relates the electron transfer matrix $T_+(E)$ to the hole transfer matrix $T_-(-E)$ of the opposite energy. To find how the transfer matrix transforms under $P$, we can act on Eq. (S39) with $P$ from the left. Comparing to the same equation (S39) written for the opposite-energy, we find

$$\sigma_y T_+^*(E) \sigma_y = T_-(-E).$$ (S41)

For the scattering matrix, Eq. (S40), we find the relation (in both abstract and in component form)

$$-\sigma_z S_+^*(E)\sigma_z = S_-(-E), \quad r_\pm^{(\prime)}(E) = -r_\mp^{(\prime)*}(-E), \quad d_\pm^{(\prime)}(E) = d_\mp^{(\prime)*}(-E).$$ (S42)

The transfer and scattering matrices are also constrained by the symmetry of the junction Hamiltonian (3) (main text) under the combined reversal of time and magnetic field. The time-reversal operator is $T = -i\sigma_y K$, while $R_M = \sigma_z$ has the sole effect of reversing the Zeeman term, as can be checked from Eq. (3): $R_M M(x)\sigma_x R_M^{-1} = -M(x)\sigma_x$. The combined symmetry operator $TR_M = \sigma_x K$ commutes with the junction Hamiltonian. The effect of the combined symmetry on Eq. (S38) is

$$T_\pm[M(x)] = \sigma_x T_\pm^*[M(x)]\sigma_x.$$ (S43)

[The square brackets remind us that $T_\pm$ is a functional of $M(x)$.] By using Eq. (S43) in Eq. (S40), we find that the left and right transmission amplitudes are equal,

$$d_\pm = d'_\pm.$$ (S44)

Note also that the sole action of time-reversal on Eq. (S38) yields

$$T_\pm[M(x)] = \sigma_y T_\pm^*[-M(x)]\sigma_y.$$ (S45)

Combining Eqs. (S43) and (S45), we see that the diagonal elements of the transfer matrices are even functionals of $M(x)$, while the off-diagonal elements are odd. For the scattering matrices, Eq. (S40), it implies that the transmission amplitudes $d_\pm^{(\prime)}$ are even in $M(x)$, while the reflection amplitudes $r_\pm^{(\prime)}$ are odd.

## C. Perturbative treatment of $M(x)$

Because the helical edge states of a TI are protected by time-reversal symmetry from elastic backscattering, the junction is reflectionless in absence of $M(x)$. In this section we consider a weakly reflective S-TI-S junction and derive the scattering matrix to first order in $M(x)$. As explained at the end of the previous section, we expect the reflection amplitude to vanish linearly, $r_\pm^{(l)} \sim M$, for small $M$, while the transmission amplitude remains non-zero, $|d_\pm^{(l)}| = 1$. The corrections to the latter will be of order $M^2$ and we neglect them in this section.

To first order in $M$, we can expand the transfer matrix as $T(x_j, x_{j-1}) = T^{(0)}(x_j, x_{j-1}) + \delta T(x_j, x_{j-1})$. Here $x_j$, $x_{j-1}$ are *any* two points in the junction. The zeroth-order-in-$M$ transfer matrix $T^{(0)}$ is diagonal in spin-indices. From Eq. (S32) we find

$$T^{(0)}(x_j, x_{j-1}) = \exp \frac{i}{v} \int_{x_{j-1}}^{x_j} [E\tau_z + \mu - V(x')]\sigma_z dx'.$$ (S46)

The first order correction $\delta T(x_j, x_{j-1}) \sim M$ can be written in a simple form for *close enough* points $x_j$, $x_{j-1}$. Assuming $M(x)$ does not vary significantly over a short interval $[x_j, x_{j-1}]$, we find from Eq. (S38) to lowest order in $(x_j - x_{j-1})$

$$\delta T(x_j, x_{j-1}) = \frac{1}{v}\tau_z \sigma_y M(x_j)(x_j - x_{j-1}).$$ (S47)

Consider next the full transfer matrix $T(L, 0)$ of the junction. It can be decomposed as a product

$$T(L, 0) = \lim_{N\to\infty} T(x_N, x_{N-1})\dots T(x_j, x_{j-1})\dots T(x_1, x_0), \quad x_j = jL/N.$$ (S48)

Expanding Eq. (S48) to first order in $\delta T$ yields

$$T(L, 0) = T^{(0)}(L, 0) + T^{(1)}(L, 0),$$ (S49)

where

$$T^{(1)}(L, 0) = \lim_{N\to\infty} \sum_{j=1}^{N} T^{(0)}(x_N, x_j)\delta T(x_j, x_{j-1})T^{(0)}(x_{j-1}, x_0)$$

$$= \frac{1}{v}\int_0^L dx T^{(0)}(L, x)\tau_z \sigma_y M(x)T^{(0)}(x, 0).$$ (S50)

Using Eqs. (S40), (S46), (S49), and (S50), we find the scattering amplitudes to lowest order in $M$,

$$d_\pm = d'_\pm = \exp \frac{i}{v}\int_0^L [E \pm (\mu - V(x'))]dx',$$ (S51)

and

$$r_\pm = -i\frac{1}{v}\int_0^L dx M(x) \exp \frac{2i}{v}\int_0^x [E \pm (\mu - V(x'))]dx',$$
$$r'_\pm = -i\frac{1}{v}\int_0^L dx M(x) \exp \frac{2i}{v}\int_x^L [E \pm (\mu - V(x'))]dx'.$$ (S52)

Alternatively, the right reflection amplitude $r'_\pm$ can be written as

$$r'_\pm = -i\frac{1}{v}\int_0^L dx M(L - x) \exp \frac{2i}{v}\int_0^x [E \pm (\mu - V(L - x'))]dx'.$$ (S53)

This shows that $r = r'$ if both $M$ and $V$ are reflection symmetric, $M(x) = M(L - x)$ and $V(x) = V(L - x)$.

In a simple example of $M(x) = M\Theta(x)\Theta(L - x)$, $V = 0$, and $M \ll v/L$, we find

$$r_\pm = r'_\pm = M\frac{1 - e^{\frac{2i}{v}[E\pm\mu]L}}{2[E \pm \mu]},$$ (S54)

as quoted in the main text below Eq. (10).

The equations (S51)-(S52) are the main results of this section, and will be put into use in Sec. SM6.

## SM5.  SYMMETRIES OF THE HAMILTONIAN (3) OF THE MAIN TEXT

In this section we discuss the symmetries of the model (3) of the main text which lead to degeneracies at special values of the phase difference $\varphi_0$. We will not consider zero energy excitations in this section – their existence was discussed below Eq. (S9). The Hamiltonian (3) *anti*commutes with $P = \sigma_y \tau_y K$ ($K$ denotes complex conjugation). This particle-hole symmetry does not lead to any degeneracies in the excitation spectrum, and we shall not discuss it any further here.

We start by considering a reflectionless junction, $M(x) = 0$ [see remark below Eq. (S11)]. At $M(x) = 0$ the system is time-reversal symmetric (TRS) for real values of the order parameter $\Delta_0 e^{i\varphi_0}$, i.e., at $\varphi_0 = 0, \pi$. At these values of the phase, the time-reversal symmetry operator, $T = i\sigma_y K$, commutes with the Hamiltonian (3) of the main text. Since $T^2 = -1$, this leads to Kramers degeneracies at $\varphi_0 = 0$ or $\pi$, regardless of $\mu$ or $V(x)$ (see Sec. SM6 A).

A non-zero $M(x)$ breaks time-reversal symmetry and for generic $\mu$ and $V(x)$ this leads to the opening of gaps at $\varphi_0 = 0$ and $\pi$. Curiously, the degeneracies at $\varphi_0 = \pi$ are preserved if $\mu = V(x) = 0$ and $M(x)$ is reflection symmetric about the junction center, $M(x) = M(L - x)$. The degeneracy exists because there are two operators, $S_1 = \sigma_x \tau_x$ and $S_2 = \tau_y R$, that commute with $\mathcal{H}^{(0)}$ at $\varphi_0 = \pi$. Here $R$ is reflection operator, $Rf(x) = f(L - x)$. Since these two operators $S_1$ and $S_2$ do not commute there is a degeneracy at $\varphi_0 = \pi$. This degeneracy can be lifted by $\mu \tau_z \neq 0$, see Eq. (11) of the main text and the associated footnote.

## SM6.  WAVE FUNCTIONS AND SPECTRUM IN A LONG ALMOST BALLISTIC JUNCTION, $L \gg \xi$

In this section we derive the sub-gap energy spectrum and the transition matrix elements in a weakly reflecting ($|r_\pm^{(\prime)}| \ll 1$), long ($L \gg \xi$) junction. We shall focus on $\varphi_0 \approx \pi$ where, as discussed in the main text, the energy spectrum of the high Andreev levels ($E_n > E_{\mathcal{M}}$) is characterized by avoided crossings.

We start, in Sec. SM6 A, by deriving the spectrum and wave functions for a reflectionless time-reversal symmetric junction (allowing for scalar disorder). In Secs. SM6 B and SM6 C we derive the lowest order ($\propto r$) backscattering corrections near $\varphi_0 = \pi$ to wave functions and spectrum of states $n \geq 1$ and $n = 0$ (Majorana, $\mathcal{M}$), respectively. In Sec. SM6 D we show how to use the derived wave functions to find the squared transition matrix elements, Eq. (12) of the main text. Finally, in Sec. SM6 E, we calculate the correction $\propto r^2$ to the slope $dE_{\mathcal{M}}/d\varphi_0$ of the lowest level.

### A.  Spectrum and wave functions in a reflectionless junction, $r_\pm = r_\pm' = 0$

For a reflectionless junction, Eq. (S16) factors into

$$\left( e^{2i\alpha(E)} e^{-i\varphi_0} - d_+(E)d_-(E) \right) \left( e^{2i\alpha(E)} e^{i\varphi_0} - d_+(E)d_-(E) \right) = 0, \quad e^{i\alpha(E)} = \frac{E + i\sqrt{\Delta_0^2 - E^2}}{\Delta_0}, \tag{S55}$$

which has degenerate solutions at $\varphi_0 = 0$ and $\pi$ (as expected from TRS, see Sec. SM5). Using Eq. (S51) we have $d_+ d_- = \exp 2iEL/v$ and Eq. (S55) takes the form [S6]

$$EL/v - \arccos \frac{E}{\Delta_0} = \pm \frac{\varphi_0}{2} + \pi m, \quad m \in \mathbb{Z}. \tag{S56}$$

Assuming $E \ll \Delta_0$, we can expand $\arccos E/\Delta_0 \approx \pi/2 - E/\Delta_0$ and find a solution

$$E_n^{(0)}(\varphi_0) = \frac{v}{2(L + \xi)} \pi \left( n + 1 - \frac{1 + (-1)^n}{2} \right) + (-1)^n \frac{v}{2(L + \xi)} |\varphi_0 - \pi|, \quad n \in \mathbb{Z}. \tag{S57}$$

Our assumption is valid when $L \gg \xi = v/\Delta_0$, and $n \ll L/\xi$. The superscript in $E^{(0)}$ indicates zeroth order in reflection amplitudes. The factor $1 + (-1)^n$ vanishes for odd $n$ and is there to make sure that levels $n$ and $n + 1$ (with odd $n$) are degenerate at $\varphi_0 = \pi$. The states with odd (even) $n$ have positive (negative) slope $dE_n^{(0)}/d\varphi_0$ for $\varphi_0 < \pi$. In the reflectionless junction we consider here, the slope changes sign discontinuously at $\varphi_0 = \pi$. Disorder in the junction makes this switching continuous as we see below in Sec. SM6 B.

The wave functions for Andreev levels $n \geq 1$ in the reflectionless limit can be found from Eq. (S17e), by employing Eq. (S16) when carefully taking the limit $r \to 0$. Consider first $\varphi_0 < \pi$. For even $n$ we have $d_+ d_- = e^{2i\alpha} e^{-i\varphi_0}$ and

therefore $u_{n\uparrow}(L) = 0$ directly from Eq. (S17e). For the odd $n$ we find $v_{n\uparrow}(L) = 0$. For any $\varphi_0$, we find at $x = L$,

$$\Phi_{n \text{ even}}^{(0)}(L) = \Theta(\pi - \varphi_0) v_{n\uparrow}(L) \begin{pmatrix} 0 \\ e^{-i\alpha(E_n^{(0)})} e^{i\varphi_0} \\ 0 \\ 1 \end{pmatrix} + \Theta(\varphi_0 - \pi) u_{n\uparrow}(L) \begin{pmatrix} 1 \\ 0 \\ e^{-i\alpha(E_n^{(0)})} e^{-i\varphi_0} \\ 0 \end{pmatrix}, \quad n \text{ even}, \quad \text{(S58a)}$$

$$\Phi_{n \text{ odd}}^{(0)}(L) = \Theta(\pi - \varphi_0) u_{n\uparrow}(L) \begin{pmatrix} 1 \\ 0 \\ e^{-i\alpha(E_n^{(0)})} e^{-i\varphi_0} \\ 0 \end{pmatrix} + \Theta(\varphi_0 - \pi) v_{n\uparrow}(L) \begin{pmatrix} 0 \\ e^{-i\alpha(E_n^{(0)})} e^{i\varphi_0} \\ 0 \\ 1 \end{pmatrix}, \quad n \text{ odd}. \quad \text{(S58b)}$$

The wave functions $\Phi_n^{(0)}(x)$ at any position can be constructed from the boundary values $\Phi_n^{(0)}(L)$ by using the transfer matrix introduced in Sec. SM4. In the proximitized regions they decay exponentially according to Eq. (S37). Inside the junction the wave functions are extended,

$$\Phi_{n \text{ even}}^{(0)}(x) = \Theta(\pi - \varphi_0) v_{n\uparrow}(L) \begin{pmatrix} 0 \\ e^{-i\alpha(E_n^{(0)})} e^{i\varphi_0} e^{-\frac{i}{v} \int_L^x [E_n^{(0)} + \mu - V(x')] dx'} \\ 0 \\ e^{-\frac{i}{v} \int_L^x [-E_n^{(0)} + \mu - V(x')] dx'} \end{pmatrix}$$
$$+ \Theta(\varphi_0 - \pi) u_{n\uparrow}(L) \begin{pmatrix} e^{\frac{i}{v} \int_L^x [E_n^{(0)} + \mu - V(x')] dx'} \\ 0 \\ e^{-i\alpha(E_n^{(0)})} e^{-i\varphi_0} e^{\frac{i}{v} \int_L^x [-E_n^{(0)} + \mu - V(x')] dx'} \\ 0 \end{pmatrix}, \quad 0 < x < L, \quad \text{(S59)}$$

here given for even $n$ ($\Phi_n^{(0)}$ with odd $n$ has similar form). Since the Hamiltonian $\mathcal{H}^{(0)}$, Eq. (3) in the main text, commutes with $S_z$, the states $\Phi_n$ are also $S_z$-eigenstates. Normalizing the wave function, $\int dx \Phi_n^{(0)\dagger}(x) \Phi_n^{(0)}(x) = 1$, leads to

$$u_{n\uparrow}(L) = v_{n\uparrow}(L) = \frac{1}{\sqrt{2(L + \kappa_n^{-1})}}, \quad \kappa_n = \frac{1}{v}\sqrt{\Delta_0^2 - E_n^2}. \quad \text{(S60)}$$

When $E \ll \Delta_0$, we have $\kappa^{-1} \approx \xi$, and $u_{n\uparrow}(L) = v_{n\uparrow}(L) = 1/\sqrt{2(L + \xi)}$.

Consider next the lowest Andreev level, $n = 0$. Assuming $E \ll \Delta_0$, we can use Eq. (S57) with $n = 0$ and take the positive solution $E_0^{(0)} = E_{\mathcal{M}}^{(0)} > 0$. We find

$$E_{\mathcal{M}}^{(0)}(\varphi_0) = \frac{v}{2(L + \xi)} |\pi - \varphi_0|. \quad \text{(S61)}$$

The wave function $\Phi_{\mathcal{M}}$ of the state $E_{\mathcal{M}}^{(0)} \geq 0$ is discontinuous at $\varphi_0 = \pi$:

$$\Phi_{\mathcal{M}}^{(0)}(L) = \frac{1}{\sqrt{2(L + \xi)}} \left[ \Theta(\pi - \varphi_0) \begin{pmatrix} 0 \\ e^{-i\alpha(E_{\mathcal{M}}^{(0)})} e^{i\varphi_0} \\ 0 \\ 1 \end{pmatrix} - \Theta(\varphi_0 - \pi) \begin{pmatrix} 1 \\ 0 \\ e^{-i\alpha(E_{\mathcal{M}}^{(0)})} e^{-i\varphi_0} \\ 0 \end{pmatrix} \right]. \quad \text{(S62)}$$

Note that in Eq. (S62) the wave function at $\varphi_0 < \pi$ is related to that at $\varphi_0 > \pi$ by a particle-hole transformation $P = \sigma_y \tau_y K$ and simultaneous $E_{\mathcal{M}}^{(0)} \to -E_{\mathcal{M}}^{(0)}$.

### B. Backscattering corrections to spectrum and wave functions of Andreev levels $n \geq 1$

Let us next consider the correction to spectrum near $\varphi_0 = \pi$ due to a small reflection amplitude $r$. The spectrum equation (S16) can be written as

$$\left(1 - \frac{d_+ d_- e^{-2i\alpha}}{|d_-|^2} e^{-i\varphi_0}\right)\left(1 - \frac{d_+ d_- e^{-2i\alpha}}{|d_+|^2} e^{i\varphi_0}\right) - e^{-2i\alpha}\left(d_+ d_- r_-'^* - r_+' e^{i\varphi_0}\right)\left(d_+ d_- r_+'^* - r_-' e^{-i\varphi_0}\right) = 0. \quad \text{(S63)}$$

(Here the scattering amplitudes and $\alpha$ are of course energy-dependent; we shall mostly omit the argument $E$ from these quantities in this section.)

In deriving Eq. (S63) we used the identities (derived from unitarity of the scattering matrix) $r_\pm = -d_\pm r'^*_\pm/d'^*_\pm$ and $d_\pm d'_\pm - r_\pm r'_\pm = d_\pm/d'^*_\pm$, and $d_\pm = d'_\pm$, Eq. (S44). In the last factors in Eq. (S63) we used $|d_\pm| = 1$, neglecting terms of order $r^4$. We shall perform a double expansion of Eq. (S63) in both $r$ and $\varphi_0 - \pi$ to find the backscattering corrections to spectrum near the degeneracy point of the clean limit. In the first two factors in Eq. (S63) we can replace $|d_\pm|^{-2} \approx 1$ since the corrections $\propto r^2$ to the norms do not contribute at zeroth order in $\varphi_0 - \pi$. In the last two factors which are both $\propto r'$, we use the zeroth order in $\varphi_0 - \pi$ expressions, $e^{\pm i\varphi_0} = -1$ and $d_+ d_- = -e^{2i\alpha(E^{(0)}(\pi))}$. The former is obtained from Eq. (S55) and $E^{(0)}(\pi)$ is the energy of the reflectionless junction at $\varphi_0 = \pi$, see Eq. (S57). Upon doing these approximations and expanding the first term in $\varphi_0 - \pi$, Eq. (S63) becomes

$$\left(1 + d_+ d_- e^{-2i\alpha}\right)^2 + (\varphi_0 - \pi)^2 + \left|r'_- - e^{2i\alpha} r'^*_+\right|^2_{E=E^{(0)}(\pi)} = 0. \tag{S64}$$

Let us denote $\delta E \sim \max(r, |\varphi_0 - \pi|)$ the small correction to the energy, $E = E^{(0)}(\pi) + \delta E$. From Eq. (S51) we have $d_+ d_- = e^{2i(E^{(0)}(\pi) + \delta E)L/v}$ up to negligible corrections $\propto r^2$ to the norms $|d_\pm|$. Using $\alpha(E) \approx \alpha(E^{(0)}(\pi)) - \delta E/\Delta_0$ and the zeroth order solution $e^{2i E^{(0)}(\pi)L/v} = -e^{2i\alpha(E^{(0)}(\pi))}$ we find $d_+ d_- e^{-2i\alpha(E)} = -e^{2i\delta E(L+\xi)/v}$. Expanding the latter to first order in $\delta E$ we can solve for it in Eq. (S64):

$$\delta E = \pm \frac{v}{2(L+\xi)} \sqrt{(\varphi_0 - \pi)^2 + \left|r'_- - e^{2i\alpha} r'^*_+\right|^2_{E=E^{(0)}_n(\pi)}}. \tag{S65}$$

The spectrum near the avoided crossing is then (see Fig. 2a in the main text)

$$E_n(\varphi_0) = \frac{v}{2(L+\xi)} \pi \left(n + 1 - \frac{1+(-1)^n}{2}\right) + (-1)^n \frac{v}{2(L+\xi)} \sqrt{(\varphi_0 - \pi)^2 + (\delta\varphi_n)^2}, \tag{S66}$$

and reproduces Eq. (S57) in the reflectionless limit. Equation (S66) is Eq. (10) in the main text. The width of the avoided crossing region is,

$$\delta\varphi_n = \left|r'^*_- - e^{-2i\alpha} r'_+\right|_{E=E^{(0)}_n(\pi)}. \tag{S67}$$

The above equation is written in terms of the right reflection amplitudes $r'_\pm$. By using the identity $r'_\pm = -d_\pm r^*_\pm/d^*_\pm$, and $d_+ d_- = -e^{2i\alpha}$ at energy $E^{(0)}(\pi)$ one can show that Eq. (S67) can be also written in terms of the left reflection matrices, $\delta\varphi_n = \left|r_+ - e^{2i\alpha} r^*_-\right|_{E=E^{(0)}_n(\pi)}$. To lowest order in $E_n/\Delta_0$, we have $e^{2i\alpha} = -1$ and we get $\delta\varphi_n = \left|r_+ + r^*_-\right|_{E=E^{(0)}_n(\pi)}$ introduced in the main text. In a simple example of $M(x) = M\Theta(x)\Theta(L - x)$, $V = 0$, and $M, \mu \ll v/L$, we find from Eq. (S54)

$$\left(r_+ + r^*_-\right)_{E=E^{(0)}_n(\pi)} = -2i \frac{M}{v/L} \frac{\mu}{E^{(0)}_n(\pi)}. \tag{S68}$$

Using the above equation in the expression for $\delta\varphi_n$ in the limit $E_n \ll \Delta_0$ leads to Eq. (11) of the main text.

Next, we derive the wave functions $\Phi_n(L)$ near the avoided crossing. From Eqs. (S17e)–(S17g) we find

$$\Phi_n(L) = \begin{pmatrix} u_{n\uparrow}(L) \\ e^{-i\alpha} e^{i\varphi_0} v_{n\uparrow}(L) \\ e^{-i\alpha} e^{-i\varphi_0} u_{n\uparrow}(L) \\ v_{n\uparrow}(L) \end{pmatrix}, \quad u_{n\uparrow}(L) = \frac{-d_+ r'^*_-/d^*_- + r'_+ e^{i\varphi_0}}{e^{i\alpha} - d_+(1/d^*_-) e^{-i\alpha} e^{-i\varphi_0}} v_{n\uparrow}(L). \tag{S69}$$

As before, we expand $u_{n\uparrow}(L)$ in small $\varphi_0 - \pi$ and $r'_\pm$ by inserting $d_+ d_- e^{-2i\alpha} = -e^{2i\delta E(L+\xi)/v}$. We find

$$u_{n\uparrow}(L) \approx -i e^{i\alpha(E^{(0)}_n(\pi))} \frac{(r'^*_- - e^{-2i\alpha} r'_+)_{E=E^{(0)}_n(\pi)}}{(\varphi_0 - \pi) - (-1)^n \sqrt{(\varphi_0 - \pi)^2 + (\delta\varphi_n)^2}} v_{n\uparrow}(L) \tag{S70}$$

up to corrections of order $(\delta E)^2$. The correction to $\Phi^\dagger_n(L)\Phi_n(L)$ by $r$ is of second order, and beyond our accuracy. Therefore, the normalization factor $v_{n\uparrow}(L)$ is determined by requiring $\Phi^\dagger_n(L)\Phi_n(L) = 1/(L+\xi)$, see Sec. SM6 A. This yields

$$v_{n\uparrow}(L) = \frac{1}{\sqrt{2(L+\xi)}} \frac{1}{\sqrt{2}} \sqrt{1 - (-1)^n \frac{(\varphi_0 - \pi)}{\sqrt{(\varphi_0 - \pi)^2 + (\delta\varphi_n)^2}}}. \tag{S71}$$

Inserting (S71) into Eq. (S70), we find

$$u_{n\uparrow}(L) = ie^{i\alpha}(-1)^n e^{i\vartheta_n} \frac{1}{\sqrt{2(L+\xi)}} \frac{1}{\sqrt{2}} \sqrt{1 + (-1)^n \frac{(\varphi_0 - \pi)}{\sqrt{(\varphi_0 - \pi)^2 + (\delta\varphi_n)^2}}} \,, \tag{S72}$$

where $\vartheta_n = \arg(r_-'^* - e^{-2i\alpha} r_+')$ and all the energies are evaluated at $E = E_n^{(0)}(\pi)$. It is convenient to introduce

$$\sin\theta_n = \frac{1}{\sqrt{2}} \sqrt{1 + \frac{(\varphi_0 - \pi)}{\sqrt{(\varphi_0 - \pi)^2 + (\delta\varphi_n)^2}}}, \quad \cos\theta_n = \frac{1}{\sqrt{2}} \sqrt{1 - \frac{(\varphi_0 - \pi)}{\sqrt{(\varphi_0 - \pi)^2 + (\delta\varphi_n)^2}}} \,, \tag{S73}$$

so that

$$v_{n\,\mathrm{odd}\uparrow}(L) = \frac{\sin\theta_n}{\sqrt{2(L+\xi)}}, \qquad\qquad u_{n\,\mathrm{odd}\uparrow}(L) = -ie^{i\alpha(E_n^{(0)}(\pi))} e^{i\vartheta_n} \frac{\cos\theta_n}{\sqrt{2(L+\xi)}} \,, \tag{S74}$$

$$v_{n\,\mathrm{even}\uparrow}(L) = \frac{\cos\theta_n}{\sqrt{2(L+\xi)}}, \qquad\qquad u_{n\,\mathrm{even}\uparrow}(L) = ie^{i\alpha(E_n^{(0)}(\pi))} e^{i\vartheta_n} \frac{\sin\theta_n}{\sqrt{2(L+\xi)}} \,. \tag{S75}$$

Finally, in terms of the wave functions (S58a), (S58b) of the reflectionless junction, the perturbed wave functions (S69) near the avoided crossing are

$$\Phi_{n\,\mathrm{even}}^{(0)}(L) = \frac{\cos\theta_n}{\sqrt{2(L+\xi)}} \begin{pmatrix} 0 \\ e^{-i\alpha(E_n)} e^{i\varphi_0} \\ 0 \\ 1 \end{pmatrix} + ie^{i\alpha(E_n^{(0)}(\pi))} e^{i\vartheta_n} \frac{\sin\theta_n}{\sqrt{2(L+\xi)}} \begin{pmatrix} 1 \\ 0 \\ e^{-i\alpha(E_n)} e^{-i\varphi_0} \\ 0 \end{pmatrix}, \quad n \text{ even}, \tag{S76a}$$

$$\Phi_{n\,\mathrm{odd}}^{(0)}(L) = -ie^{i\alpha(E_n^{(0)}(\pi))} e^{i\vartheta_n} \frac{\cos\theta_n}{\sqrt{2(L+\xi)}} \begin{pmatrix} 1 \\ 0 \\ e^{-i\alpha(E_n)} e^{-i\varphi_0} \\ 0 \end{pmatrix} + \frac{\sin\theta_n}{\sqrt{2(L+\xi)}} \begin{pmatrix} 0 \\ e^{-i\alpha(E_n)} e^{i\varphi_0} \\ 0 \\ 1 \end{pmatrix}, \quad n \text{ odd}. \tag{S76b}$$

### C. Backscattering correction to the wave function $\Phi_{\mathcal{M}}$ of Andreev level $n = 0$

In this section we will calculate the backscattering correction $\delta\Phi_{\mathcal{M}}$ to the wave function of the Majorana doublet ($n = 0$ state). The presence of a $E_{\mathcal{M}} = 0$ state at $\varphi_0 = \pi$ is not removed by any $r$. An illustration of this is that $\delta\varphi_{n=0} = 0$ in Eq. (S67), as can be verified by using the properties $r_+'(0) = -r_-'^*(0)$ [see Eq. (S42)] and $e^{2i\alpha(0)} = -1$ in that equation. Unlike the crossing of the levels ($E_{\mathcal{M}}, -E_{\mathcal{M}}$), the degeneracy at $\varphi_0 = 0$ of the pair ($E_{\mathcal{M}}, E_1$) is lifted by $r$. A finite value of $r$ modifies also the slope $dE_{\mathcal{M}}/d\varphi_0$ at $\varphi_0 = \pi$ by a small correction $\propto r^2$, calculated in Sec. SM6 E. Since we are interested here only in the first-order corrections, we will neglect the modification of $E_{\mathcal{M}}$, and focus on the first-order correction to the wave function.

Let us first consider phase $\varphi_0 < \pi$. Then the unperturbed wave function (S62) has $u_{\mathcal{M}\uparrow}^{(0)}(L) = 0$. From the second equation in Eq. (S69), we have to lowest order in $r_{\pm}'$

$$u_{\mathcal{M}\uparrow}(L) = \frac{-d_+ d_- r_-'^* + r_+' e^{i\varphi_0}}{e^{i\alpha} - d_+ d_- e^{-i\alpha} e^{-i\varphi_0}} v_{\mathcal{M}\uparrow}(L) \,. \tag{S77}$$

Both the numerator and denominator in the above equation vanish at $\varphi_0 = \pi$ as can be checked by using the properties at $E_{\mathcal{M}} = 0$: $r_+'(0) = -r_-'^*(0)$, $e^{i\alpha(0)} = i$, and $d_+ d_- = 1$. Taking the limit $\varphi_0 \to \pi$ yields (we use $d/d\varphi_0 = \partial_{\varphi_0} E \partial_E + \partial_{\varphi_0}$)

$$u_{\mathcal{M}\uparrow}(L) = i \frac{2\partial_{\varphi_0} E_{\mathcal{M}}(-r_+' d_+^{-1} \partial_E d_+ + \partial_E r_+') + ir_+'}{-2\partial_{\varphi_0} E_{\mathcal{M}}(d_+^{-1} \partial_E d_+ - i\partial_E \alpha(E)) + i} v_{\mathcal{M}\uparrow}(L) \,. \tag{S78}$$

To zeroth order in $r$ the transmission amplitude satisfies $d_+^{-1} \partial_E d_+ = iL/v$, see Eq. (S51). Also, for $\varphi_0 \to \pi^-$ we have $-2\partial_{\varphi_0} E_{\mathcal{M}} = v/(L+\xi)$. Finally, using $\partial_E \alpha(E) \approx -\xi/v$ and the condition $r_+' = -r_+^* d_+^2$ following from unitarity of $S_+$, we get

$$u_{\mathcal{M}\uparrow}(L) = \frac{v}{2L} d_+^2 \partial_E r_+^* \frac{1}{\sqrt{2(L+\xi)}}, \quad \varphi_0 < \pi \,, \tag{S79}$$

where we also used $v_{\mathcal{M}\uparrow}(L) = 1/\sqrt{2(L+\xi)}$. The corrected wave function at $\varphi_0 < \pi$ is

$$\Phi_{\mathcal{M}}(L) = \frac{1}{\sqrt{2(L+\xi)}} \left[ \begin{pmatrix} 0 \\ e^{-i\alpha(E_{\mathcal{M}})}e^{i\varphi_0} \\ 0 \\ 1 \end{pmatrix} + \delta\Phi_{\mathcal{M}} \begin{pmatrix} 1 \\ 0 \\ e^{-i\alpha(E_{\mathcal{M}})}e^{-i\varphi_0} \\ 0 \end{pmatrix} \right], \quad \varphi_0 < \pi, \tag{S80}$$

where

$$\delta\Phi_{\mathcal{M}} = \frac{v}{2L} d_+^2 \partial_E r_+^*, \tag{S81}$$

and the scattering amplitudes and their derivatives are evaluated at $E = 0$. As seen in Eq. (12) of the main text, the correction $\delta\Phi_{\mathcal{M}}$ modifies the transition matrix elements slightly.

The wave function at $\varphi_0 > \pi$ can be found by solving for $v_{0\uparrow}(L)$ in Eq. (S17e) with the help of (S16), and then doing a calculation similar to the one above. A more convenient way is to use the particle-hole transformation $P$ [see remark below Eq. (S62)] on Eq. (S80). We find

$$\Phi_{\mathcal{M}}(L) = \frac{1}{\sqrt{2(L+\xi)}} \Theta(\pi - \varphi_0) \left[ \begin{pmatrix} 0 \\ e^{-i\alpha(E_{\mathcal{M}})}e^{i\varphi_0} \\ 0 \\ 1 \end{pmatrix} + \delta\Phi_{\mathcal{M}} \begin{pmatrix} 1 \\ 0 \\ e^{-i\alpha(E_{\mathcal{M}})}e^{-i\varphi_0} \\ 0 \end{pmatrix} \right]$$
$$- \frac{1}{\sqrt{2(L+\xi)}} \Theta(\varphi_0 - \pi) \left[ \begin{pmatrix} 1 \\ 0 \\ e^{-i\alpha(E_{\mathcal{M}})}e^{-i\varphi_0} \\ 0 \end{pmatrix} + \delta\Phi_{\mathcal{M}}^* \begin{pmatrix} 0 \\ e^{-i\alpha(E_{\mathcal{M}})}e^{i\varphi_0} \\ 0 \\ 1 \end{pmatrix} \right]. \tag{S82}$$

In the simple example of $M(x) = M\Theta(x)\Theta(L-x)$, $V = 0$, and $M \ll v/L$, we find from Eqs. (S51), (S54), (S81)

$$\delta\Phi_{\mathcal{M}} = \frac{ML}{v} \left( \frac{1}{2} + i\frac{1}{3}\frac{L\mu}{v} \right). \tag{S83}$$

This result was quoted below Eq. (12) in the main text.

The example used in the main text in the context of a NW was a $\delta$-impurity at the NW-S interface, $M(x) = V(x) = u_0\delta(x-L)$. The scattering matrices for a more generic $\delta$-impurity at $x = x_0$ are calculated in Sec. SM8. In our specific example we find to lowest order in $u_0/v$,

$$\delta\Phi_{\mathcal{M}} = \frac{u_0}{v} \left( 1 - i\frac{u_0}{v} \right). \tag{S84}$$

### D. The transition matrix elements

In the main text we saw that the oscillator strengths appearing in the admittance $Y$ are given by the matrix elements $\mathcal{H}^{(1)}_{\pm\mathcal{M};n}$ of the perturbation $\partial_{\varphi_0}\mathcal{H}^{(0)}$. From Eqs. (3) and (6) of the main text we can read off $\partial_{\varphi_0}\mathcal{H}^{(0)}(x) = -\Theta(x-L)\Delta_0[\tau_x \sin\varphi_0 + \tau_y \cos\varphi_0]$ and

$$\mathcal{H}^{(1)}_{\mathcal{M};n} = -\Delta_0 \int_L^\infty dx\, \Phi_{\mathcal{M}}^\dagger(x)[\tau_x \sin\varphi_0 + \tau_y \cos\varphi_0]\Phi_n(x). \tag{S85}$$

The integration domain covers only the proximitized region, where the wave functions decay exponentially from their boundary values at $x = L$, see Eq. (S37). Using the wave functions at $x = L$ near $\varphi_0 = \pi$ [derived above, Eqs. (S76a), (S76b) and (S82)] in Eq. (S85), we find the transition matrix elements to lowest order in reflection amplitudes and $|\varphi_0 - \pi|$. Squaring the matrix elements leads to Eq. (12) of the main text.

### E. Backscattering correction to the slope $dE_{\mathcal{M}}/d\varphi_0$ of Andreev level $n = 0$

In this section we calculate the terms denoted $\mathcal{O}(|r_+|^2)$ in Eq. (S11). To access these second order corrections, we need the transmission amplitudes $d_\pm$ to that order. They can be obtained by expanding the transfer matrix to second order in $M$. One finds

$$d_\pm(E) = d'_\pm(E) = (1 - \frac{1}{2}|r_\pm(E)|^2)d_\pm^{(0)}(E), \tag{S86}$$

where $r_\pm(E)$ is given by Eq. (S52) and $d_\pm^{(0)}(E) = \exp \frac{i}{v} \int_0^L [E \pm (\mu - V(x'))] dx'$ from Eq. (S51). The form (S86) satisfies the requirement $|d_\pm|^2 + |r_\pm|^2 = 1$ to second order in $r$.

We can now insert Eq. (S86) in the denominator in Eq. (S10) and keep terms up to second order in $r$. By using the properties $\partial_E d_\pm^{(0)} = \frac{iL}{v} d_\pm^{(0)}$, $|d_\pm^{(0)}| = 1$, and $r_-^*(E) = -r_+(E)$, we find

$$\left| \partial_E d_+ + i d_+ \Delta_0^{-1} \right|^2 - \left| \partial_E (d_+ r_-^*) \right|^2 \approx \left( \frac{L + \xi}{v} \right)^2 (1 - |r_+|^2) - \left| \partial_E r_+ - \frac{iL}{v} r_+ \right|^2, \tag{S87}$$

at $E = 0$, and

$$\partial_{\varphi_0} E_{\mathcal{M}}(\pi) \approx \pm \frac{1}{2} \frac{v}{L + \xi} \left( 1 - \frac{1}{2} |r_+|^2 + \frac{1}{2} \frac{v^2 \left| \partial_E r_+ - \frac{iL}{v} r_+ \right|^2}{(L + \xi)^2} \right)_{E=0}. \tag{S88}$$

This is the equation referred to in the footnote 30 of the main text. In the simple example with $M(x) = M\Theta(x)\Theta(L - x)$, $V = 0$, and $M, \mu \ll v/L$, we have $\partial_E r_+ = \frac{iL}{v} r_+$ to lowest order in $\mu/(v/L)$. The second term in (S88) is then negligible, and we find

$$\partial_{\varphi_0} E_{\mathcal{M}}(\pi) \approx \pm \frac{1}{2} \frac{v}{L + \xi} (1 - \frac{1}{2} |r_+(0)|^2) = \pm \frac{1}{2} \frac{v}{L + \xi} |d_+(0)|. \tag{S89}$$

Note that $|r_+(0)| = |r_-(0)|$. In the second equality we used Eq. (S86).

## SM7. EXTENSION TO NANOWIRE JUNCTION

The effective Hamiltonian $H^{(0)} = \int dx \Psi^\dagger \mathcal{H}^{(0)} \Psi / 2$ of the nanowire junction is [S7, S8]

$$\mathcal{H}^{(0)}(x) = -\tau_z \frac{\partial_x^2}{2m} - i\alpha \tau_z \tilde{\sigma}_z \partial_x - \mu \tau_z + U(x) \tau_z + B\tilde{\sigma}_x + \Delta(x) [\tau_x \cos\varphi(x) - \tau_y \sin\varphi(x)], \tag{S90}$$

$$\Delta(x) = \Delta[\Theta(-x) + \Theta(x - L)], \ \varphi(x) = \varphi_0 \Theta(x - L). \tag{S91}$$

Here $m$ is the effective mass of the electrons, $\alpha$ is the SO coupling constant, $U(x)$ is a scalar disorder potential, $B$ is the Zeeman energy, and $\Delta$ is the proximity-induced gap. The Pauli matrices $\tilde{\sigma}$ act on the spin-space and the Nambu spinor is $\Psi^\dagger = (\psi_\uparrow^\dagger, \psi_\downarrow^\dagger, \psi_\downarrow, -\psi_\uparrow)$. Coupling to a weak microwave source is treated as in the main text.

We will next show that at low energies the spin-orbit coupled nanowire is very similar to the TI edge, and its low-energy Hamiltonian is identical to Eq. (3) of the main text. For illustration, we shall focus on the limit of large Zeeman energy, $B \gg m\alpha^2, E$, and set $\mu = U(x) = 0$. Near the Fermi points $k \approx \pm k_Z = \pm \sqrt{2mB}$, the effective Hamiltonian is $H_{\text{eff}}^{(0)} = \int dx \phi^\dagger \mathcal{H}_{\text{eff}}^{(0)} \phi / 2$ with [S9]

$$\mathcal{H}_{\text{eff}}^{(0)}(x) = -iv\tau_z \sigma_z \partial_x + \Delta_0(x) [\tau_x \cos\varphi(x) - \tau_y \sin\varphi(x)], \tag{S92}$$

where $v = k_Z/m$, $\Delta_0(x) = \Delta(x) 2m\alpha/k_Z$, and $\phi^\dagger = (R^\dagger, L^\dagger, L, -R)$ is the slow mode field operator. The Pauli matrices $\sigma_{x,y,z}$ and $\tau_{x,y,z}$ act on the respective spaces of left and right movers, and particles and holes. The above Hamiltonian is derived from Eq. (S90) by using the projections [S9]

$$\psi_\uparrow(x) \approx -\gamma_- e^{ik_Z x} R(x) - \gamma_+ e^{-ik_Z x} L(x), \tag{S93}$$
$$\psi_\downarrow(x) \approx \gamma_+ e^{ik_Z x} R(x) + \gamma_- e^{-ik_Z x} L(x),$$

and assuming slowly varying operators $L(x)$ and $R(x)$. Here $\gamma_\pm = (1 \pm m\alpha/k_Z)/\sqrt{2}$.

The low-energy Hamiltonian (S92) has the same form as the Hamiltonian (3) of the main text with $\mu = V = M = 0$, and therefore the bound state spectrum of the clean ($U = 0$) nanowire junction is the same as that of a reflectionless S-TI-S system (with of course different Dirac velocity $v$ and pairing gap $\Delta_0$).

Above we derived Eq. (S92) for a clean nanowire with no potential disorder $U(x)$. For a simple model of disorder, consider a $\delta$-impurity at position $x = x_0$ in the junction,

$$H_{\text{imp}} = \sum_{\tilde{\sigma} = \uparrow, \downarrow} \int dx u_0 \delta(x - x_0) \psi_{\tilde{\sigma}}^\dagger \psi_{\tilde{\sigma}}, \tag{S94}$$

or $U(x) = u_0 \delta(x - x_0)$ in Eq. (S90). Using Eq. (S93), the projection of $H_{\text{imp}}$ to low energies is

$$
\begin{aligned}
H_{\text{imp}} = & \ u_0 [\gamma_-^2 + \gamma_+^2] \left[ R^\dagger(x_0) R(x_0) + L^\dagger(x_0) L(x_0) \right] \\
& + u_0 2\gamma_+ \gamma_- \left[ e^{-2ik_Z x_0} R^\dagger(x_0) L(x_0) + e^{2ik_Z x_0} L^\dagger(x_0) R(x_0) \right] .
\end{aligned}
\tag{S95}
$$

The phases in the second term can be gauged away by setting $R(x) \to e^{-2ik_Z x_0} R(x)$. The projected $H_{\text{imp}}$ leads to an additional term of the form

$$
\mathcal{H}_{\text{imp}}^{(0)} = [v_0 \tau_z + m_0 \sigma_x] \delta(x - x_0) ,
\tag{S96}
$$

in Eq. (S92). Here

$$
\begin{aligned}
v_0 &= u_0 [\gamma_-^2 + \gamma_+^2] , \\
m_0 &= 2 u_0 \gamma_+ \gamma_- ,
\end{aligned}
\tag{S97}
$$

are the strengths of forward and backscattering due to $H_{\text{imp}}$. The effect of $\mathcal{H}_{\text{imp}}^{(0)}$ on the sub-gap spectrum is studied in Sec. SM8 below. Generally, the presence of an impurity leads to lifting of degeneracies at $\varphi_0 = 0, \pi$ (without removing $E_\mathcal{M}(\pi) = 0$ of course). In particular, we find that an impurity at the NW-S interface, $x_0 = L$, results in a width $\delta\varphi_\pi \propto m_0/v$ of the avoided crossing at $\varphi_0 = \pi$. This result was quoted in the main text, Eq. (13), and will be derived in the next section.

## SM8. SINGLE $\delta$-IMPURITY – SCATTERING AMPLITUDES AND EQUATION FOR SPECTRUM

For a single $\delta$-impurity the transmission and reflection amplitudes can be calculated explicitly for any impurity strength. Consider an impurity

$$
M(x) = m_0 \delta(x - x_0) , \quad V(x) = v_0 \delta(x - x_0) ,
\tag{S98}
$$

where $0 \le x_0 \le L$. We have encountered such an impurity at $x_0 = L$ in the main text in the paragraph of Eq. (13) (see also Sec. SM7). In this section we will study the effect of the impurity on the spectrum of Andreev levels and derive Eq. (13).

For the above $\delta$-impurity the junction transfer matrix can be written as

$$
T(L, 0) = T(L, x_0) e^{\frac{-i}{v} \tau_z \sigma_z (v_0 \tau_z + m_0 \sigma_x)} T(x_0, 0) .
\tag{S99}
$$

where $T(L, x_0)$ and $T(x_0, 0)$ correspond to reflectionless propagation,

$$
T(L, x_0) = e^{\frac{i}{v}(L - x_0)[E\tau_z + \mu]\sigma_z} , \quad T(x_0, 0) = e^{\frac{i}{v} x_0 [E\tau_z + \mu]\sigma_z} .
\tag{S100}
$$

Inserting Eq. (S100) into Eq. (S99), we find in the space of left and right movers

$$
T(L, 0) = \begin{pmatrix} e^{\frac{i}{v}L[E\tau_z + \mu]} \frac{1}{z^*} & -i\tau_z e^{\frac{i}{v}(L - 2x_0)[E\tau_z + \mu]} \frac{w}{z} \\ i\tau_z e^{-\frac{i}{v}(L - 2x_0)[E\tau_z + \mu]} \frac{w}{z} & e^{-\frac{i}{v}L[E\tau_z + \mu]} \frac{1}{z} \end{pmatrix} .
\tag{S101}
$$

The complex numbers $z$, $w$ satisfy $|w|^2 + |z|^2 = 1$. Explicitly they are

$$
z = \left( \cos \frac{1}{v} \sqrt{v_0^2 - m_0^2} + i \frac{v_0}{\sqrt{v_0^2 - m_0^2}} \sin \frac{1}{v} \sqrt{v_0^2 - m_0^2} \right)^{-1} ,
\tag{S102a}
$$

$$
w = z \frac{m_0}{\sqrt{v_0^2 - m_0^2}} \sin \frac{1}{v} \sqrt{v_0^2 - m_0^2} .
\tag{S102b}
$$

Using Eq. (S40) and above $T(L, 0)$, the normal state scattering matrices are

$$
\begin{pmatrix} r_+ & d'_+ \\ d_+ & r'_+ \end{pmatrix} = \begin{pmatrix} -i e^{\frac{i}{v} 2x_0 [E + \mu]} w & e^{\frac{i}{v} L[E + \mu]} z \\ e^{\frac{i}{v} L[E + \mu]} z & -i e^{2\frac{i}{v}(L - x_0)[E + \mu]} w \end{pmatrix} ,
\tag{S103a}
$$

$$
\begin{pmatrix} r_- & d'_- \\ d_- & r'_- \end{pmatrix} = \begin{pmatrix} -i e^{\frac{i}{v} 2x_0 [E - \mu]} w^* & e^{\frac{i}{v} L[E - \mu]} z^* \\ e^{\frac{i}{v} L[E - \mu]} z^* & -i e^{2\frac{i}{v}(L - x_0)[E - \mu]} w^* \end{pmatrix} .
\tag{S103b}
$$

We see that $|r_\pm^{(\prime)}| = |w|$ and $|d_\pm| = |z|$ are independent of energy. The above-derived transmission amplitudes satisfy

$$\frac{d_+}{d_-^*} = \frac{d_-}{d_+^*} = e^{\frac{i}{v}2LE}. \tag{S104}$$

With the above properties, the equation for spectrum, Eq. (S16), simplifies significantly. We find [S10]

$$|d|^2 \cos\varphi_0 - \cos 2[\frac{LE}{v} - \alpha(E)] - |r|^2 \cos\frac{2(L - 2x_0)E}{v} = 0, \tag{S105}$$

where we denote $|r_\pm^{(\prime)}| = |r|$ and $|d_\pm| = |d|$, and

$$|d|^2 = \frac{1}{1 + \frac{m_0^2}{v_0^2 - m_0^2}\sin^2\frac{1}{v}\sqrt{v_0^2 - m_0^2}} = 1 - |r|^2. \tag{S106}$$

If the impurity is at the interface, $x_0 = 0, L$, a gap of order $|r|(v/L)(\xi/L)$ opens for the high Andreev levels at $\varphi_0 = \pi$ (for simplicity, we assume $|r| \ll 1$ here). Equation (S67) gives the width of the resulting avoided crossing region,

$$\delta\varphi_n = 2\frac{\xi}{L}|r|\frac{E_n^{(0)}(\pi)}{v/L}. \tag{S107}$$

If the impurity originates from a scalar impurity in a NW junction (see Sec. SM7), we can relate $|r|$ to the microscopic impurity strength $u_0$. This is done by using Eq. (S97) to express $v_0$ and $m_0$ in terms of the microscopic parameters, and using them in Eq. (S106). We find that when $u_0 \ll v\sqrt{B/m\alpha^2}$, the reflection coefficient becomes $|r|^2 = Z^2/(4 + Z^2)$ with $Z = 2m_0/v \approx 2u_0/v$.

For weak backscattering ($u_0 \ll v$) we insert $|r| \approx |u_0|/v \ll 1$ in the expression for $\delta\varphi_n$ to find Eq. (13) of the main text. The phase $\vartheta_n$ in this case is $e^{i\vartheta_n} = (r_-'^* - e^{-2i\alpha}r_+')_{E=E_n^{(0)}(\pi)} = (u_0/|u_0|)(1 - iu_0/v)$.

Finally, we note that an impurity in the middle of the junction does not lift any of the degeneracies at $\varphi_0 = \pi$. This can be seen from Eq. (S105) where setting $x_0 = L/2$ yields

$$\cos 2[\frac{LE}{v} - \alpha(E)] = |d|^2\cos\varphi_0 - |r|^2. \tag{S108}$$

Setting $\varphi_0 = \pi$ and using $|r|^2 = 1 - |d|^2$ makes the right-hand-side independent of $|r|$. The spectrum at $\varphi_0 = \pi$ is therefore identical to that of a reflectionless junction and has degeneracies. These degeneracies arise because of destructive interference between the scattered electron and hole optical paths when the impurity is exactly in the middle of the junction. This result is in agreement with Ref. [S10].

---

[S1] $N_R = \sum_{\sigma=\uparrow,\downarrow}\int_L^{L+a}dx\,\psi_\sigma^\dagger(x)\psi_\sigma(x)$ where $a \gg \xi$ is the width of the tunneling region.

[S2] A. C. Potter and P. A. Lee, Phys. Rev. B 83, 184520 (2011).

[S3] C. Wang, Y. Y. Gao, I. M. Pop, U. Vool, C. Axline, T. Brecht, R. W. Heeres, L. Frunzio, M. H. Devoret, G. Catelani, L. I. Glazman, and R. J. Schoelkopf, Nature Communications 5, 5836 (2014).

[S4] C. W. J. Beenakker, Phys. Rev. Lett. 67, 3836 (1991).

[S5] L. Fu and C. L. Kane, Phys. Rev. B 79, 161408 (2009).

[S6] C. W. J. Beenakker, D. I. Pikulin, T. Hyart, H. Schomerus, and J. P. Dahlhaus, Phys. Rev. Lett. 110, 017003 (2013).

[S7] R. M. Lutchyn, J. D. Sau, and S. Das Sarma, Phys. Rev. Lett. 105, 077001 (2010).

[S8] Y. Oreg, G. Refael, and F. von Oppen, Phys. Rev. Lett. 105, 177002 (2010).

[S9] J. Klinovaja and D. Loss, Phys. Rev. B 86, 085408 (2012).

[S10] S.-f. Zhang, W. Zhu, and Q.-f. Sun, Journal of Physics: Condensed Matter 25, 295301 (2013).



\end{document}